\documentclass[traditabstract]{aa}
\usepackage{amsmath}
\usepackage{amssymb}
\usepackage{graphicx}
\usepackage{cite}
\usepackage{color} 
\usepackage{mathrsfs}
\usepackage{mathtools}
\usepackage{bm}
\usepackage{graphicx}
\usepackage{url}
\usepackage{bm}
\usepackage{enumitem}
\usepackage{multirow}
\usepackage{booktabs}
\usepackage{array}
\usepackage{colortbl}
\usepackage[square,numbers,sort&compress]{natbib}

\usepackage[labelfont=bf,labelsep=period,justification=raggedright]{caption}
\makeatletter
\renewcommand{\@biblabel}[1]{\quad#1.}
\makeatother

\date{}

\newcommand{\shaderow}{\rowcolor[gray]{.85}}
\newcommand{\about}{\raise.17ex\hbox{$\scriptstyle\sim$}}
\newcommand{\BF}{\mathrm{BF}}
\newcommand{\PIP}{\mathrm{PIP}}
\newcommand{\Var}{\mathrm{Var}}
\newcommand{\lb}{\bm{[}}
\newcommand{\rb}{\bm{]}}

\newcommand{\smfrac}[2]{{\textstyle\frac{#1}{#2}}}
\newcommand{\half}{{\smfrac{1}{2}}}
\newcommand{\Data}{\mathscr{D}}
\newcommand{\sfbf}[1]{\textsf{\textbf{#1}}}

\begin{document}

\title{Integrated analysis of variants and pathways in
    genome-wide association studies using polygenic models of
    disease}

\titlerunning{Integrated analysis of variants and pathways in
    genome-wide association studies}

\authorrunning{Carbonetto and Stephens}

   \author{Peter Carbonetto \inst{1} \and
           Matthew Stephens \inst{2}}

   \institute{Dept. of Human Genetics,
              University of Chicago 
              \and
              Depts. of Statistics and Human Genetics,
              University of Chicago}
 
\maketitle

\section*{Abstract}

Many common diseases are highly polygenic, modulated by a large number
genetic factors with small effects on susceptibility to disease. These
small effects are difficult to map reliably in genetic association
studies. To address this problem, researchers have developed methods
that aggregate information over sets of related genes, such as
biological pathways, to identify gene sets that are {\em enriched} for
genetic variants associated with disease. However, these methods fail
to answer a key question: which genes and genetic variants are
associated with disease risk? We develop a method based on sparse
multiple regression that simultaneously identifies enriched pathways,
and prioritizes the variants within these pathways, to locate
additional variants associated with disease susceptibility. A central
feature of our approach is an estimate of the strength of enrichment,
which yields a coherent way to prioritize variants in enriched
pathways. We illustrate the benefits of our approach in a genome-wide
association study of Crohn's disease with $\about$440,000 genetic
variants genotyped for $\about$4700 study subjects. We obtain strong
support for enrichment of {\sf\em IL-12}, {\sf\em IL-23} and other
cytokine signaling pathways. Furthermore, prioritizing variants in
these enriched pathways yields support for additional
disease-association variants, all of which have been independently
reported in other case-control studies for Crohn's disease.

\section*{Author Summary}

Genome-wide association studies have helped locate genes that
contribute to common diseases. The analysis of these studies is
typically straightforward: systematically test each genetic variation
in isolation whether it is correlated with susceptibility to disease.
This approach often works well to identify commonly occurring variants
with moderate effects on disease risk, but the effects of many
variants are so small they fail to show statistically significant
correlations. This is a concern because many common diseases are
modulated by a large number of genetic factors with small effects on
disease risk. An alternative strategy is to examine groups of
variants, such as variants sharing a common biological pathway, and to
test whether each group is “enriched” for disease correlations. This
can be a more effective approach to identify genetic factors relevant
to disease. However, it does not indicate which individual genes or
variants are associated with disease, which remains an important
question. To address this limitation, we describe a statistical
framework that integrates enrichment analysis with disease-correlation
tests for variants. We illustrate this approach in a case-control
study for Crohn's disease. We show that our approach uncovers
disease-susceptibility genes that are not identified in conventional
analyses of the same data.

\section*{Introduction}

By surveying genetic variation throughout the genome, and
systematically searching for variants correlated with disease
phenotypes, genome-wide association studies (GWAS) have led to the
discovery of genes and genetic loci that were not previously suspected
of playing a role in complex diseases \cite{altshuler-2008,
  frazer-2009, mccarthy-2008, pearson-manolio-2008}. For example, the
discovery of disease-correlated variants in GWAS of Crohn's disease, a
common form of inflammatory bowel disease, has helped draw links to
genes that regulate autophagy and innate immune responses
\cite{abraham-2009c, barrett-2008, franke-2010, khor-2011,
  stappenbeck-2011, vanlimbergen-2009}.

In most analyses of GWAS, genetic associations are identified by
testing each marker one at a time for association with disease.
Additional clues about genetic variants, such as whether they are
coding, exonic, or lie near the transcription start site of a gene,
are usually considered {\em post hoc}, rather than incorporated into
the statistical analysis. While such ``hypothesis-free'' analyses have
successfully identified variants associated with disease, they have
important shortcomings. Many commonly occurring diseases are believed
to be modulated by a large number of genetic factors, each which have
a small effect on disease risk, so they may be difficult to identify
in a GWAS \cite{eichler-2010, manolio-2009, rioux-abbas-2005,
  ropers-2007}. This problem is compounded, according to some
predictions \cite{bodmer-bonilla-2008, eichler-2010, manolio-2009,
  pritchard-2001}, by the low prevalence of many alleles conferring
risk to complex diseases. Motivated by these shortcomings, researchers
have developed ``pathway analysis'' approaches to GWAS
\cite{ballard-2010, braun-2011, ramanan-2012, wang-2010,
  yaspan-2011}. These methods are grounded on the theory that complex
disease arises from the accumulation of genetic effects acting within
common biological pathways \cite{cantor-2010, hartwell-2004,
  hirschhorn-2009, schadt-2009}. A main goal of pathway analysis is to
identify pathways that are {\em enriched} for disease---that is,
groups of related genes that are more likely to harbour
disease-associated variants compared to arbitrary regions of the
genome. Pathway analysis can improve power to uncover genetic factors
relevant to disease because identifying the accumulation of small
genetic effects acting in a common pathway is often easier than
mapping the individual genes within the pathway that contribute to
disease susceptibility.

A limitation of analyses that identify enriched pathways is that they
do not provide feedback about associated variants within these
pathways; identifying an enriched set of genes does not indicate which
variants are associated with the disease, or even which genes harbour
such variants. Yet discovery of associated variants and genes is a
primary motivation for GWAS, and an important step toward
understanding the biology of disease and, ultimately, developing
effective medical therapies. In principle, identifying enriched
pathways should help us locate variants associated with disease,
because knowing that a variant lies in or near a gene in an enriched
pathway makes it a better disease-association candidate. Therefore,
prioritizing these variants---say, by slightly relaxing thresholds of
significance---should yield a greater number of reproducible
associations for a given rate of false positives. But it is unclear
how to implement this in practice: to what extent can we relax
significance thresholds while keeping the rate of false positives at
an acceptable level?

To address this problem, we develop a model-based approach for joint
analysis of pathways and genetic variants, in which we interpret
enrichment as a model parameter. The enrichment parameter quantifies
the increase in the probability that each variant in the pathway is
associated with disease risk. By jointly analyzing variants and
pathways, our method adjusts association evidence in light of
estimated pathway enrichments---sometimes called {\em pathway} or {\em
  gene prioritization} \cite{aerts-2006, baranzini-2009, cantor-2010,
  chen-2011, franke-2006, lage-2007, raychaudhuri-2009, saccone-2008,
  tranchevent-2011, wu-2008}---and, simultaneously, adjusts enrichment
estimates to reflect evidence of associations in pathways.

Our approach builds on statistical methods for simultaneous mapping of
genetic variants in GWAS \cite{bottolo-richardson-2010, bvs,
  guan-stephens-2011, he-lin-2011, hoggart-2008, logsdon, segura-2012,
  yi-2008, wu-2009}. In contrast to single-marker regression
approaches, these methods model susceptibility to disease by the
combined effect of multiple variants, and use sparse multivariate
regression techniques to fit multi-marker ({\em i.e.} polygenic)
models to the data. By adopting a multi-marker disease model,
estimating enrichment effectively reduces to counting, inside and
outside each candidate pathway, variants associated with
disease---more precisely, the variants that are included in the
polygenic disease model. Our approach to combining multi-marker
modeling with pathway analysis offers several benefits. First,
compared to single-marker approaches, multi-marker modeling improves
power to detect genetic associations \cite{guan-stephens-2011,
  platt-2010, segura-2012}. Second, unlike many pathway analysis
methods that test for enrichment of significant SNPs or genes in a
pathway \cite{wang-2010}, we have no need to select a significance
threshold for {\em p}-values; instead, we use the association signal
across all genes to assess enrichment. Third, by analyzing variants
simultaneously, we avoid exaggerating evidence for enrichment from
associated variants that are correlated with each other ({\em i.e.}
in linkage disequilibrium), while still allowing multiple independent
association signals near a gene to contribute evidence for
enrichment. And fourth, quantifying enrichment within this framework
naturally gives us feedback about associations within enriched
pathways, potentially leading to discovery of novel genetic loci
underlying disease.

Though we focus on incorporating pathways (and, more broadly,
biologically related gene sets) into analysis of GWAS, our methods
also apply to other types of genome annotations. In this respect, our
work is related to other model-based approaches for estimating
enrichment of genome-wide association signals across functionally
related genomic regions, like transcription factor binding sites
\cite{gaffney-2012, lee-2009, lewinger-2007, veyrieras-2008}. One
distinguishing feature of our approach is that we have the ability to
test for enrichment, which is important for assessing which candidate
pathways show the strongest support for enrichment. We assess support
for enrichment by framing it is as a model comparison problem. An
advantage of this approach is that we can use the same approach to
assess support for the enrichment of combinations of two or more
pathways. This is useful, for example, if a pathway relevant to
disease pathogenesis is ranked highly only after combining it with
another pathway relevant to the disease.

Another feature that distinguishes our analysis is that we use
multiple pathway databases in an attempt to interrogate pathways as
comprehensively as possible---the more pathways we consider, the
greater chance we have of drawing new connections between pathways,
genes within these pathways, and complex disease. We demonstrate how
using our approach to comprehensively interrogate pathways results in
increased evidence for enrichment, and is robust to inclusion of a
large number of irrelevant pathways. Our study includes $\about{3100}$
candidate pathways drawn from eight well-developed pathway databases
available on the Web \cite{pathguide-2006, bauer-mehren-2009,
  tranchevent-2011}.

We demonstrate our approach in a detailed analysis of a GWAS for
Crohn's disease with about $440,000$ single nucleotide polymorphisms
(SNPs) genotyped in roughly $1700$ cases and $3000$ controls. This is
a convenient study for gauging the benefits of our approach because
genetic associations have already been published based on data from
this study \cite{wtccc}, and pathway analyses of these data have found
evidence for enriched pathways \cite{ballard-2010, holmans-2009,
  peng-2010, torkamani-2008, wang-2009}. Our enrichment results
highlight the role of cytokines that modulate immune responses in
Crohn's disease, and the {\sf\em IL-12} and {\sf\em IL-23} signaling
pathways, which have been previously linked to the disease
\cite{abraham-cho-2009b, khor-2011, wang-2009}. And, by prioritizing
variants within these enriched pathways, our method identifies
disease-susceptibility candidates that are not deemed significant in
conventional analyses of the same data, including the {\sf\em STAT3}
gene, the {\sf\em IBD5} locus, and the major histocompatibility
complex (MHC) class II genes. All these genetic associations have been
independently confirmed in other studies, demonstrating that our
methods have the potential to yield novel biological insights.

\section*{Overview of statistical analysis}

Our approach builds on previous work that casts simultaneous analysis
of genetic variants as a {\em variable selection} problem---the
problem of deciding which variables (the genetic variants) to include
in a multivariate regression of the phenotype. We begin with a method
that assumes each variant is equally likely to be associated with the
phenotype \cite{bvs, guan-stephens-2011}, then we modify this
assumption to allow for enrichment of associated variants in a
pathway.
  
The data from the GWAS are the genotypes $\mathbf{X} = (x_1, \ldots,
x_n)^T$ and phenotypes $y = (y_1, \ldots, y_n)^T$ from $n$ study
participants. Here we assume the genetic markers are SNPs, and the
phenotype is disease status: patients with the disease (``cases'') are
labeled $y_i = 1$, and disease-free individuals (``controls'') are
labeled $y_i = 0$. Entries of the $n \times p$ matrix $\mathbf{X}$ are
observed minor allele counts $x_{ij} \in \{0,1,2\}$, or expectations
of these counts estimated using genotype imputation \cite{li-2009,
  marchini-2010}, for each of the $n$ samples and $p$ SNPs.

We assume an additive model of disease risk, in which the log-odds for
disease is a linear combination of the minor allele counts:
\begin{align}
\log\bigg\{\frac{p(y_i=1)}{p(y_i=0)}\bigg\} = 
\beta_0 + x_{i1}\beta_1 + \cdots + x_{ip}\beta_p.
\label{eq:logistic}
\end{align}
Under this additive model, $e^{\beta_j}$ is the {\em odds ratio,} the
multiplicative increase in odds of disease for each copy of the
minor allele at locus $j$. We do not consider dominant or recessive
effects on disease risk, but it would be straightforward to include
them; see \cite{servin-stephens-2007}. This method is also easily
adapted to quantitative traits by replacing \eqref{eq:logistic} with a
linear regression for $y$.

Although the log-odds for disease is expressed in \eqref{eq:logistic}
as a linear combination of all SNPs, we assume most SNPs $j$ have no
effect on disease risk ($\beta_j = 0$). We refer to SNPs $j$ that have
non-zero coefficients ($\beta_j \neq 0$) as being ``included'' in the
multi-marker disease model. A SNP's inclusion signals that it affects
susceptibility to disease, or that is in linkage disequilibrium with
other, possibly untyped, risk-conferring variants. Therefore, the main
goal of the analysis is to compute, for each SNP $j$, the posterior
inclusion probability, $\PIP(j) \equiv p(\beta_j \neq 0 \,|\,
\mathbf{X}, y)$. A high posterior inclusion probability is the
analogue of a small {\em p}-value in a conventional single-marker
analysis.

To obtain these posterior inclusion probabilities, we must first
specify a prior. A standard assumption, and the assumption made in
previous approaches \cite{bvs, guan-stephens-2011}, is that SNPs are
equally likely to be associated with the phenotype {\em a priori};
that is, $\pi_j \equiv p(\beta_j \neq 0)$ is the same for all SNPs.
  
To model enrichment of associations within a pathway, we modify this
prior. Precisely, the prior inclusion probability for SNP $j$ depends
on whether or not it is assigned to the enriched pathway:
\begin{align}
\log_{10}\bigg(\frac{\pi_j}{1-\pi_j}\bigg) = \theta_0 + a_j\theta. 
\label{eq:pathway-prior}
\end{align}
The pathway indicators $a_j$ keep track of which SNPs are assigned to
the enriched pathway: $a_j = 1$ when SNP $j$ is assigned to the
enriched pathway, otherwise $a_j = 0$. (In brief, a SNP is assigned to
a pathway if it is near a gene in the pathway; see Methods.) We refer
to $\theta_0$ as the {\em genome-wide log-odds,} since it reflects the
overall proportion of SNPs that are included in the multi-marker
disease model. (More precisely, it is the proportion for SNPs not
assigned to the pathway, but this is usually most SNPs.) We refer to
$\theta$ as the {\em enrichment parameter} because it corresponds to
the increase in probability {\rm (on the log-odds scale)} that a SNP
assigned to the pathway is included in the model. For example,
$\theta_0 = -5$ and $\theta = 2$ indicates that 1 out of every 10,000
SNPs outside the pathway is included in the multi-marker model, but
for SNPs assigned to the pathway, 1 out of every 100 are included. If
$\theta=0$, this reduces to the standard prior assumption made by
previous methods. We expect $\theta$ to be zero, or close to zero, for
most pathways.

To assess evidence for enrichment of a candidate pathway with
indicators $a = (a_1, \ldots, a_p)$, we compute a {\em Bayes factor}
\cite{kass-raftery-1995, stephens-balding-2009}:
\begin{equation}
\BF(a) = 
\frac{p(y \,|\, \mathbf{X}, a, \theta > 0)}
     {p(y \,|\, \mathbf{X}, \theta = 0)}.
\label{eq:BF}
\end{equation}
This Bayes factor (BF) is the ratio of likelihoods under two models,
the model in which the candidate pathway is enriched for SNPs included
in the multi-marker regression ($\theta > 0$), and the null model that
no pathways are enriched ($\theta = 0$). A larger BF implies stronger
evidence for enrichment.  We compute each BF by averaging, or {\em
  integrating}, over the unknown parameters, and over multi-marker
models with different combinations of SNPs, using appropriate prior
distributions for $\theta_0$ and $\theta$ (see Methods). 

We use the same approach to test for joint enrichment of multiple
candidate pathways. We compute $\BF(a)$ as before (eq.~\ref{eq:BF}),
except that we set $a_j$ to 1 whenever SNP $j$ is assigned to at least
one of the pathways. In this case, $\theta$ represents the increased
rate of associations among SNPs assigned to one or more of the
pathways. This is equivalent to assuming that all enriched pathways
have the same level of enrichment. It would be possible to relax this
assumption, but at the cost of complicating the analysis, so we
restrict ourselves to a single enrichment parameter.

To assess evidence for association of individual variants with the
phenotype, we compute $\PIP(j)$ for each variant $j$. These posterior
probabilities depend on which pathways are enriched, and on the
strength of enrichment $\theta$, because these factors affect the
prior probabilities $\pi_j$, which in turn affect the posterior
probabilities $\PIP(j)$ following Bayes' rule. (In practice, we account
for uncertainty in $\theta_0$ and $\theta$ when calculating the
posterior probabilities by averaging over $\theta_0$ and $\theta$; see
Methods.)  Since enrichment leads to higher prior inclusion
probabilities for SNPs in the enriched pathway, an association that is
not identified by a conventional genome-wide analysis, perhaps because
the allele appears infrequently in the population, or because the
nearby functional polymorphism has only a small effect on disease
risk, may become a strong candidate in light of its presence in an
enriched pathway. Because we estimate $\theta$ from the data, the
extent to which we prioritize variants in enriched pathways is
determined by the data. In this way, our framework integrates the
problem of identifying enriched pathways with the problem of
prioritizing variants within enriched pathways.

\section*{Results}

We illustrate our methods through an extended example---analysis of
genome-wide marker data from the WTCCC Crohn's disease study
\cite{wtccc}. After steps to ensure data quality, the data consist of
$\about{440,000}$ SNPs genotyped for 1748 cases and 2938
controls. (See Methods for details.) Crohn's disease is well suited to
illustrating the benefits of pathway analysis because many pathways
related to immune function and inflammatory response have been
characterized. Additionally, genetic associations have been published
based on data from this study \cite{wtccc}, and have been replicated
in a follow-up study \cite{parkes-2007}. Beyond these association
analyses, enrichment analyses of these data have provided further
support for links between Crohn's disease and pathways related to
adaptive and innate immunity \cite{ballard-2010, holmans-2009,
  peng-2010, torkamani-2008, wang-2009}.

We have three main goals in presenting this case study: first, to
explain how to use and interpret the BFs, PIPs, and other statistics
relevant to joint analysis of variants and pathways; second, to
highlight the features, and limitations, of our approach; and third,
to examine the hypothesis that we can gain additional insights into
Crohn's disease by reassessing the evidence for associations between
variants and disease in light of pathway enrichment findings.

Our analysis proceeds in three stages. First, we compute a BF for each
candidate pathway, and use these Bayes factors to rank the pathways.
Second, based on this ranking, we assess evidence for models in which
two or more pathways are enriched.  Third, we investigate whether the
most compelling pathway enrichments can help us locate Crohn's disease
associations beyond those identified in analyses that ignore
information about pathways.

\subsection*{Bayes factors for enriched pathways}

\begin{table*}[t]
\caption{\bf Pathways, and groups of related pathways, that exhibit
    the strongest evidence for enrichment of Crohn's disease
    associations, as measured by their Bayes factors}
{\sf
\begin{tabular}{@{}m{2.25in}@{\;\;\;}c@{\;\;\;}c@{\;\;\;}c@{\;\;\;}r@{\;\;\;}c@{}}
& & number of & Bayes \\[-0.15em]
enriched pathway & database & genes/SNPs & factor & 
$\bm{\bar{\theta}_0}$ & enrichment $\bm{\bar{\theta}}$ \\ \toprule[0.5pt]
Cytokine signaling in immune system & React. (BS) & 225/6711 & 
$\mathsf{7.9\!\times\!10^5}$ & $\mathsf{-4.1}$ & 1.96 (1.25--2.50) \\
IL23-mediated signaling events & PID (PC) & 66/2218 & 
$\mathsf{9.1\!\times\!10^3}$ & $\mathsf{-3.9}$ & 2.01 (1.50--2.50) \\
IL12-mediated signaling events & PID (PC) & 111/3641 & 
$\mathsf{5.4\!\times\!10^3}$ & $\mathsf{-4.0}$ & 1.90 (1.25--2.50) \\
Immune system & React. (BS) & 755/20,959 & $\mathsf{1.3\!\times\!10^3}$ 
& $\mathsf{-4.1}$ & 1.44 (0.75--2.00) \\
Signaling by interleukins & React. (BS) & 111/3678 & 769 & $\mathsf{-3.9}$ & 
1.70 (1.25--2.25) \\
Interferon gamma signaling & React. (BS) & 73/2290 & 700 & $\mathsf{-3.8}$ & 
1.73 (1.25--2.25) \\
Immune system & React. (PC) & 529/15,074 & 692 & $\mathsf{-4.1}$ & 
1.49 (0.75--2.00) \\
Interferon signaling & React. (BS) & 111/2965 & 491 & $\mathsf{-3.8}$ & 
1.68 (1.25--2.25) \\
Cytokine signaling in immune system & React. (PC) & 193/6194 & 292 & 
$\mathsf{-3.9}$ & 1.55 (1.00--2.25) \\
Signaling events mediated by TC-PTP & PID (PC) & 92/3044 & 219 & 
$\mathsf{-3.9}$ & 1.62 (1.00--2.25) \\
\raggedright TAK1 activates NF-$\kappa$B by phosphory- lation and activation 
of IKKs complex & React. (BS) & 24/844 & 181 & $\mathsf{-3.8}$ & 
2.00 (1.25--2.50) \\
\raggedright Selective expression of chemokine receptors during 
T-cell polarization & BioCarta & 29/880 & 139 & $\mathsf{-3.8}$ 
& 2.01 (1.25--2.75) \\
IL27-mediated signaling events & PID (PC, BS) & 26/796 & 130 & $\mathsf{-3.8}$ 
& 2.00 (1.25--2.50) \\
IL23-mediated signaling events & PID (BS) & 37/1252 & 119 & $\mathsf{-3.8}$ & 
1.89 (1.25--2.50) \\
CXCR4-mediated signaling events & PID (PC) & 190/6733 & 118 & $\mathsf{-3.9}$ 
& 1.43 (0.75--2.00) \\
\raggedright Activated TAK1 mediates p38 MAPK activation
 & React. (BS) & 17/535 & 116 & $\mathsf{-3.8}$ & 2.07 (1.25--2.75) \\
TCR signaling in na\"ive CD4+ T cells & PID (PC) & 133/4809 & 103 & 
$\mathsf{-3.9}$ & 1.47 (0.75--2.00) \\
\raggedright JNK phosphorylation and activation mediated by activated 
human TAK1 & React. (BS) & 16/559 & 102 & $\mathsf{-3.8}$ & 2.03 (1.25--2.75) 
\end{tabular}}
\begin{flushleft}
The table includes all gene sets with $\BF>100$.  Abbreviations used
in this table: React. = Reactome \cite{reactome-2011}, PID = NCI
Nature Pathway Interaction Database \cite{pid-2009}, BS = NCBI
BioSystems \cite{biosystems-2010}, PC = Pathway Commons
\cite{pc-2011}; $\bar{\theta}_0=$ posterior mean of genome-wide
log-odds $\theta_0$ given that pathway is enriched; $\bar{\theta}=$
posterior mean of enrichment $\theta$ (and 95\% credible interval)
given that pathway is enriched. The credible interval is the smallest
interval about the posterior mean that contains $\theta$ with 95\%
posterior probability. It is calculated to the nearest 0.25 using a
numerical approximation (see Supplementary Methods). Note that these
numbers may not be reproduced exactly in an independent analysis using
the same method, due to stochasticity in our approximate
computations. However, only slight deviations from these numbers are
expected.
\end{flushleft}
\label{table:ranking} 
\end{table*}

To rank the 3158 candidate pathways by their evidence for enrichment,
we compute, for each pathway, a Bayes Factor that measures support for
enrichment relative to the null hypothesis. All candidate pathways
have been curated by domain experts, or are based on experimental
evidence in other organisms and inferred via gene homology. To be as
comprehensive as possible, we draw pathways from a variety of publicly
accessible collections, and we include all pathways, even those that
are unlikely to be relevant to Crohn's disease based on current
understanding of disease pathogenesis. To help make our results
replicable, we document the pathway databases used, and the steps
taken to compile gene sets from these pathway data. (See
Table~\ref{table:pathways} for the list of pathway databases used; see
Supplementary Results for statistics on gene sets and gene coverage
from these databases; see Methods for retrieval and processing of
pathway data, and assignment of SNPs to pathways.)

Many of the pathways in the databases are arranged hierarchically---we
include all elements of the hierarchy in our analysis. Elements in
upper levels of the hierarchy refer to groups of pathways with shared
attributes or a common function. Some groups have a broad definition,
such as ``immune system'' in Reactome, which includes pathways
involved in adaptive and innate immune response. (Hereafter, we use
the term ``pathway'' to refer to a set of biologically related genes.)
Enrichment of a broad group of genes is unlikely to provide novel
insights into disease pathogenesis. However, a key step in our
analysis is to re-interrogate SNPs for association in light of
inferred enrichments. Thus, enrichment of a broad physiological target
like ``immune system'' can be useful if subsequent re-interrogation
reveals associations that were not significant in a conventional
analysis.

We find that the vast majority of pathways show little or no evidence
for enrichment; of the 3518 candidate pathways, 2850 (90\%) have $\BF
< 1$, and an additional 233 pathways (7\%) have BFs between 1 and
10. Table~\ref{table:ranking} shows the 18 pathways with $\BF >
100$. The cutoff at 100, as with any cutoff, is somewhat arbitrary. We
discuss this issue and interpretation of the BFs below.

Several gene sets in Table~\ref{table:ranking} are subsets of one
another (refer to Fig.~\ref{fig:hierarchy} for relationships among
these pathways in the Reactome and PID hierarchies). For example,
``immune system'' overlaps with eight other pathways in the table,
including cytokine signaling. Several pathways appear in the table
twice because the NCBI BioSystems and Pathway Commons databases
sometimes disagree about the set of genes assigned to a pathway.
These discrepancies can have a substantial impact on the results. For
example, the BF for the Pathway Commons version of cytokine signaling
is smaller than the BioSystems version by a factor of roughly 1000,
due primarily to the lack of inclusion of {\sf\em NOD2} and MHC genes
that contribute to the association signal. Conversely, the BF for the
Pathway Commons version of IL-23 signaling is about 80 times larger
than the BioSystems version because the former includes the
NF-$\kappa$B pathway, and this pathway contains several genes that
contribute to the associational signal, notably {\sf\em
  NOD2}. (Inclusion of the NF-$\kappa$B pathway is supported by
experimental evidence \cite{cho-2006}.) These results illustrate the
benefits of a comprehensive analysis that considers pathways from
multiple sources. Also note that no pathways from HumanCyc and KEGG,
which are mainly focused on metabolic pathways, show up in
Table~\ref{table:ranking} (nor do any pathways from Cancer Cell Map,
PANTHER and WikiPathways). This points to the robustness of our
approach to inclusion of a large number of irrelevant pathways.

All the pathways in Table~\ref{table:ranking} are related in some way
to immune system function. These pathways implicate key actors in
responses to pro-inflammatory stimuli and in regulation of innate and
adaptive immunity. This includes members of the NF-$\kappa$B/Rel
family, T-cell receptors (TCRs), members of the protein tyrosine
phosphatase family (PTPs), mitogen-activated protein (MAP) kinases
such as c-Jun $\rm NH_2$-terminal kinases (JNKs), and chemokine
receptors (CXCRs) \cite{bonizzi-2004, charo-2006, dong-2002,
  pao-2007}.

In this initial ranking, three pathways stand out as having stronger
evidence for enrichment than the others: cytokine signaling in immune
system, IL23-mediated signaling events, and IL12-mediated signaling
events. The ``cytokine signaling in immune system" pathway, with $\BF =
7.9 \times 10^5$, is a collection of cytokine-driven networks that
modulate immune responses. Cytokines, a class of chemical messengers
that includes interferons (IFNs), interleukins (ILs) and tumor
necrosis factors (TNFs), have well understood roles in immune
processes. Accordingly, they exhibit a complex relationship to
autoimmunity: in addition to promoting inflammatory and immune
responses, they play an important role in suppressing immunity
\cite{oshea-2002}. Accumulating evidence points to cytokines, and the
signaling cascades initiated by these cytokines, in a range of
autoimmune disorders, including inflammatory bowel disease
\cite{godessart-kunkel-2001, stappenbeck-2011,
  zhernakova-2009}. Enrichment of ``cytokine signaling in immune
system" is consistent with this view. Even though it implicates a
broad class of genes (the BioSystems version has 225 genes), enrichment
for Crohn's disease associations is biologically plausible; immune
response involves a complex interplay among signaling pathways, so a
collection of pathways may explain the pattern of genetic associations
better than any single signal transduction pathway.

Enrichment of the IL-23 pathway for Crohn's disease associations is
consistent with its involvement in intestinal inflammation,
specifically in mediating differentiation of $\rm CD4^{+}$ Th17 cells
\cite{mcgovern-powrie-2007}. Additional findings from mouse models and
genetic association studies support its role in inflammatory bowel
disease \cite{abraham-cho-2009b, abraham-cho-2009, cho-2008,
  mcgovern-powrie-2007, stappenbeck-2011}. IL-12 is also thought to
be important for immune response and differentiation of Th17 cells,
even if many of the regulating activities previously ascribed to IL-12
are due to IL-23 instead \cite{mcgovern-powrie-2007,
  vanlimbergen-2009}. Although IL-12 and IL-23 have distinct functions
in regulation of T helper cells, their pathways have many cytokine and
cytokine receptor components in common \cite{hunter-2005}, so it may
be difficult to tease apart their roles in disease solely by looking
at enrichment in their gene sets; of the 66 genes assigned to the
IL-23 pathway, 54 are assigned to the IL-12 pathway (Pathway Commons
version).

Previous findings from genome-wide association studies have linked
autophagy genes {\sf\em ATG16L1} and {\sf\em IRGM} to Crohn's disease
\cite{hampe-2007, rioux-2007, wtccc}. Our pathway analysis does not
provide additional support for autophagy in Crohn's disease because
pathways reflecting current models of autophagy \cite{homer-2010,
  stappenbeck-2011} have not yet been incorporated, to our knowledge,
into any of the publicly available pathway databases.


Many of the pathways listed in Table~\ref{table:ranking} are related
to those identified in previous pathway analyses of Crohn's disease
\cite{ballard-2010, holmans-2009, peng-2010, torkamani-2008},
including Jak-STAT signaling \cite{peng-2010} and T cell receptor
signaling \cite{wang-2009}. Other pathways highlighted in previous
analyses show some evidence for enrichment in our analysis, but these
are eclipsed by much stronger enrichment signals
(Table~\ref{table:ranking}). For example, Wang {\em et al}
\cite{wang-2009} obtain the most evidence for enrichment of ``IL12 and
Stat4 dependent signaling in Th1 development'' ({\em p}-value = $8
\times 10^{-5}$, FDR = $0.045$) based on enrichment analysis of
BioCarta, KEGG and Gene Ontology \cite{go-2000} gene sets, whereas
this pathway showed only modest evidence for enrichment in our
analysis ($\BF = 18$) compared to the pathways in
Table~\ref{table:ranking}. (Below, we obtain greater support for
enrichment of this pathway when combined with cytokine signaling
genes.) Moveover, all except one of the pathways in
Table~\ref{table:ranking} are from databases that were not included in
\cite{wang-2009}. These results illustrate the benefits of a
comprehensive search for enrichment across multiple pathway databases.

\begin{table*}[t]
\caption{\bf Pairs of pathways with strongest evidence for joint
  enrichment of Crohn's disease associations}
{\sf
\begin{tabular}{@{}c@{}m{3in}@{\;\;\;}c@{\;\;\;}c@{\;\;\;}c@{\;\;\;\;}c@{}}
\hspace*{1em} & & & number of & Bayes \\[-0.15em]
\multicolumn{2}{@{}l@{}}{enriched pathways} & database & genes/SNPs & factor & 
$\bm{\bar{\theta}}$ \\ 
\toprule[0.5pt]
\multicolumn{2}{@{}l@{}}{cytokine signaling in immune system and $\ldots$} \\
& IL23-mediated signaling events & PID (BS) & 247/7438 & 
$\mathsf{5.5\!\times\!10^8}$ & 2.33 \\
& \raggedright Selective expression of chemokine receptors during 
T-cell polarization & BioCarta & 248/7375 & $\mathsf{4.8\!\times\!10^8}$ 
& 2.33 \\
& Th1/Th2 differentiation & BioCarta & 236/7090 & 
$\mathsf{3.6\!\times\!10^8}$ & 2.32 \\
& \raggedright $\mathsf{NO_{2}}$-dependent IL-12 pathway in NK cells &
BioCarta & 239/7143 & $\mathsf{3.2\!\times\!10^8}$ & 2.31 \\
& IL27-mediated signaling events & PID (PC, BS) & 238/7073 & 
$\mathsf{3.0\!\times\!10^8}$ & 2.30 \\
& IL12-mediated signaling events & PID (BS) & 263/7807 & 
$\mathsf{2.4\!\times\!10^8}$ & 2.30 \\
& \raggedright IL-12 and Stat4 dependent signaling pathway in Th1 
development & BioCarta & 241/7272 & $\mathsf{2.4\!\times\!10^8}$ & 2.30 \\
& IL23-mediated signaling events & PID (PC) & 266/8060 & 
$\mathsf{2.2\!\times\!10^8}$ & 2.29 
\end{tabular}}
\begin{flushleft}
This table includes every model with two enriched pathways that has a
BFs greater than $10^8$. All these combinations includes ``cytokine
signaling in immune system'' (the Biosystems version with 225
genes). Refer to Table \ref{table:ranking} for the legend. ``Number of
genes/SNPs'' gives the total number assigned to the enriched
pathways. For every enrichment hypothesis listed in the table,
$\bar{\theta}_0 = 4.4$, and the 95\% credible interval for $\theta$ to
the nearest 0.25 is (1.75--2.75). Refer to Fig.~\ref{fig:hierarchy}
for relationships among pathways in the Reactome and PID hierarchies.
\end{flushleft}
\label{table:ranking2} 
\end{table*}

Given that enrichment analyses typically proceed by computing {\em
  p}-values and assessing ``significance,'' one may reasonably ask
whether the BFs in Table~\ref{table:ranking} represent ``significant''
evidence for enrichment. Specifying an appropriate threshold for a BF
to be considered significant, however, is context-dependent, and
subjective. This is because the posterior odds for a pathway being
enriched, relative to the null hypothesis that no pathways are
enriched, is equal to the Bayes Factor times the prior odds for
enrichment, and the prior odds for each pathway depends on how
plausible it is, {\em a priori}, that the pathway is involved in
Crohn's disease pathogenesis. (Similar issues arise when specifying
significance thresholds for {\em p}-values. For example, the false
discovery rate at a given {\em p}-value threshold depends on the prior
probability of enrichment \cite{storey-2003,storey-2003b}. In
practice, however, significance thresholds of 0.05 or 0.01 are often
used without attending to this concern.) Nonetheless, we can make the
following observations. First, if we are willing to assume the
pathways in Table~\ref{table:ranking} are all equally plausible
candidates for enrichment {\em a priori}, then the ratio of the BFs
indicates the relative support for the enrichment hypotheses. For
example, if we are forced to choose between enrichment of ``cytokine
signaling in immune system'' and ``IL23-mediated signaling events,''
the data overwhelmingly favour the former by a factor of $\frac{7.9
  \times 10^5}{9.1 \times 10^3} \approx 87$. Second, even under a
``conservative'' prior for enrichment in which we expect one pathway
to be enriched among the 3158 candidates, corresponding to a prior
odds of $1/3158$, the top three pathways have BFs that are large
enough (greater than 3158) to support enrichment.

\subsection*{Assessing combinations of pathways for enrichment}

The initial ranking (Table~\ref{table:ranking}) suggests that
enrichment of a group of pathways, cytokine signaling in immune
system, offers a better fit to the pattern of Crohn's disease
associations than any one pathway. But the question remains whether
some other combination of pathways offers a better fit. A benefit of
our approach is that we can directly compare support for enrichment of
different combinations of pathways by comparing their BFs (assuming
the same prior for these hypotheses). This is because the ratio
$\BF(a)/\BF(a^{\ast})$ is the same as the Bayes factor that compares
support for the enrichment model encoded by $a$ versus the model
encoded by $a^{\ast}$. (By contrast, it is harder to make such
comparisons using {\em p}-values. For example, if $p$ is the {\em
  p}-value for testing hypothesis $a$ against the null, and $p^{\ast}$
is the {\em p}-value for testing $a^{\ast}$ against the null, it is
not clear how to compare support for $a$ and $a^{\ast}$.)

We begin by computing BFs to quantify support for enrichment of pairs
of pathways.  Since it is impractical to consider all pairs, we tackle
this in a ``greedy'' fashion by selecting combinations of pathways
based on the initial ranking. Our strategy is to select pathways with
the largest BFs (here we take IL-23, IL-12 and cytokine signaling),
and we consider each of these in combination with pathways from a
larger set of candidates (here we use the 72 pathways with $\BF >
10$). This greedy heuristic makes it feasible to evaluate many
combinations of pathways that could plausibly be jointly enriched,
though since it does not consider all combinations, there is the risk
of overlooking a combination that provides a better fit to the pattern
of associations. In total, we compute BFs for 219 pairs of pathways,
which includes 3 combinations of IL-23, IL-12 and cytokine signaling,
and $216 = 3 \times 72$ combinations of IL-23, IL-12 or cytokine
signaling with another pathway.

Table~\ref{table:ranking2} lists all models with two enriched pathways
that have $\BF>10^8$. As before, all these pathways are related to
innate and adaptive immune processes. Three of the BioCarta pathways
appearing in this table, including the IL-12 and Stat4 dependent
signaling pathway highlighted in the enrichment analysis of
\cite{wang-2009}, did not originally show up in
Table~\ref{table:ranking} because their BFs were less than 100. The
largest BF, for enrichment of IL-23 and cytokine signaling, is
unsurprising in light of the initial ranking, except that this version
of the IL-23 pathway does not include the NF-$\kappa$B pathway, which
suggests that the NF-$\kappa$B pathway does not contribute additional
evidence for enrichment once we account for enrichment of cytokine
signaling genes (14 of the 35 genes in the NF-$\kappa$B pathway,
including {\sf\em NOD2}, are also members of cytokine
signaling). Among hypotheses that do not involve cytokine signaling,
the largest BF is $5.1 \times 10^7$, corresponding to a model in which
IL-23 signaling and interferon gamma signaling are enriched.

The top BF in Table~\ref{table:ranking2} is $\about{700}$ times
greater than the largest BF in Table~\ref{table:ranking}.  This
suggests that the best model with two enriched pathways provides a
much better fit to the data than the best model with any one enriched
pathway. However, to properly interpret this result we must weigh this
increase in the BF against the relative prior plausibility of the
models, which is difficult to quantify. A naive argument using a
``conservative'' prior for any pair of pathways being enriched might
suggest a prior odds of $(1/3158)^2$, based on the conservative prior
for a single pathway we discussed above. This prior would make a
700-fold increase in the BF appear to be relatively
insignificant. However, this argument not only depends on the earlier
prior, which may be overly conservative, but also assumes independence
of enriched pathways, which seems unwise considering that many
pathways mentioned here have related roles in immunity; {\em a
  priori}, one might expect that a pathway is more likely to be
enriched when a biologically related pathway is enriched. With this in
mind, we interpret Table~\ref{table:ranking2} as providing
substantial, if short of compelling, support for the hypothesis that
two pathways are enriched for Crohn's disease associations. Perhaps a
more important question is whether these findings lead to
identification of additional loci affecting susceptibility to Crohn's
disease, a question we address in the next section.

\begin{figure*}[t]
\centering
\includegraphics[width=16cm,keepaspectratio=true]{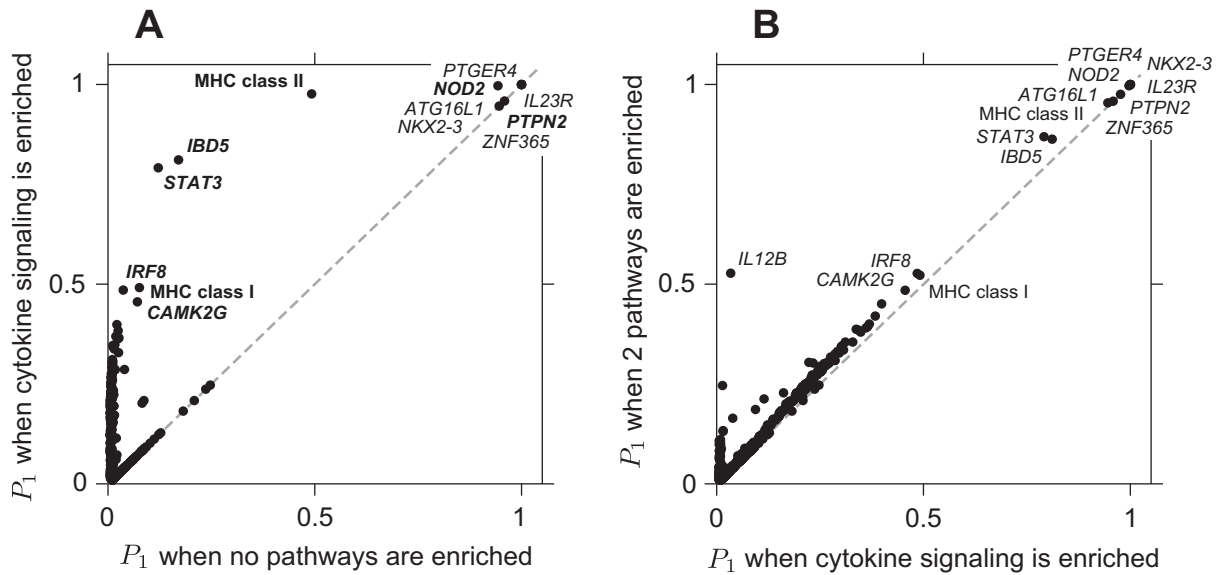}
\caption{{\bf Scatterplots showing $\bm{P_1}$, the posterior
    probability that each genomic segment contains Crohn's disease
    associations, given different hypotheses about enriched pathways.}
  Each point corresponds to a segment of the genome containing 50
  SNPs. Some of the segments are labeled by representative candidate
  genes. In panel A, genes assigned to ``cytokine signaling in immune
  system'' are written in bold.}
\label{fig:scatterplots} 
\end{figure*}

For completeness, we extend the analysis to models with three enriched
pathways. Again, following a greedy strategy, we take all pairs of
pathways with $\BF>10^8$ (Table~\ref{table:ranking2}) and combine
these pairs with individual pathways that have $\BF>100$
(Table~\ref{table:ranking}). Of the 126 resulting BFs, the largest is
$8.4 \times 10^9$, corresponding to enrichment of IL-23 signaling
(Pathway Commons), cytokine signaling (BioSystems) and ``TAK1
activates NF-$\kappa$B by phosphorylation and activation of IKKs.''
This BF is only about 3 times greater than the largest BF in
Table~\ref{table:ranking2}. Following our earlier arguments, this
result does not constitute strong support for enrichment of three
pathways.

\subsection*{Associations informed by enriched pathways}

\begin{table*}[t]
\caption{\bf Selected regions of the genome with strong evidence for
  Crohn's disease risk factors given that two pathways are enriched
  for Crohn's disease associations} 
{\sf
\begin{tabular}{@{\;}c@{\;}c@{\;\;\;\;}c@{\;\;\;}c@{\;\;\;}c@{\;\;\;}c@{\;\;\;}c@{\;\;\;}c@{\;\;\;}l@{\;\;\;}c@{\;\;\;}c@{}}
& & \raisebox{-0.5em}{\sfbf{critical}} & \multicolumn{2}{c}{\!\!$\bm{P_1}$} & 
\multicolumn{2}{c}{\!\!$\bm{P_2}$} & 
\raisebox{-0.5em}{\sfbf{candidate}} & &
\multicolumn{2}{c}{\sfbf{MAF}} \\[-0.3em]
\cmidrule(l{0.2em}r{1em}){4-5} \cmidrule(l{0.2em}r{1em}){6-7} 
\cmidrule(l{0.2em}r{0.2em}){10-11}
& \sfbf{chr.} & \sfbf{region (Mb)} & \sfbf{null} & \sfbf{alt.} & 
\sfbf{null} & \sfbf{alt.} & \sfbf{gene(s)} & \sfbf{SNP} & 
\sfbf{ctrls} & \sfbf{cases} \\
& 1p31 & 67.30--67.48 & 1.00 & 1.00 & 0.06 & 0.48 & 
\sf\em IL23R & rs11805303 & 0.318 & 0.392 \\
& 2q37 &233.92--234.27& 1.00 & 1.00 & 0.01 & 0.22 & 
\sf\em ATG16L1 & rs10210302 & 0.481 & 0.402 \\
& 5p13 & 40.32--40.66 & 1.00 & 1.00 & 0.42 & 0.42 & 
\sf\em PTGER4 & rs17234657 & 0.124 & 0.181 \\
$\ast$ & 5q23 & 129.54--132.04 & 0.18 & 0.86 & 0.02 & 0.29 & 
multiple ({\sf\em IBD5}) & rs274552 & 0.166 & 0.128 \\
$\ast$ & 6p21 & 32.3--32.92 & 0.49 & 0.98 & 0.02 & 0.31 &
MHC class II & rs9469220 & 0.519 & 0.466 \\
& 10q21 & 64.0--64.43 & 0.96 & 0.96 & 0.03 & 0.03 & 
\sf\em ZNF365 & rs10995271 & 0.386 & 0.440 \\
& 10q24 & 101.26--101.32 & 0.95 & 0.95 & 0.01 & 0.04 & 
\sf\em NKX2-3 & rs7095491 & 0.470 & 0.528 \\
& 16q12 & 49.0--49.4 & 1.00 & 1.00 & 0.11 & 0.77 & 
\sf\em NOD2 & rs17221417 & 0.287 & 0.356 \\
$\ast$ & 17q21 & 37.5--38.3 & 0.12 & 0.87 & 0.01 & 0.21 &
\sf\em STAT3 & rs744166 & 0.439 & 0.392 \\
& 18p11 & 12.76--12.91 & 0.94 & 1.00 & 0.01 & 0.20 & 
\sf\em PTPN2 & rs2542151 & 0.163 & 0.209
\end{tabular}}
\begin{flushleft}
For every region in this table, there is at least a 0.8 probability
that one or more SNPs in the region are included in the multi-marker
disease model ($P_1 \geq 0.8$) given the hypothesis that two pathways
are enriched. Rows marked with an asterisk ($\ast$) are selected only
after accounting for enriched pathways. Table columns from left to
right are: (1) chromosomal locus; (2) region most likely containing
the risk-conferring variant(s), in Megabases (Mb); (3) posterior
probability that one or more SNPs in the region are included in the
model under the null hypothesis, and (4) under the alternative
hypothesis that two pathways are enriched; (5) posterior probability
that two or more SNPs are included under the null, and (6) under the
alternative; (7) established genes in Crohn's disease pathogenesis, or
most credible genes of interest, corresponding to the locus; (8)
refSNP identifier of SNP in critical region with the largest PIP; (9)
frequency of minor allele for that SNP in cases, and (10) in
controls. The ``critical region'' at each locus is estimated by
inspecting single-SNP BFs \cite{servin-stephens-2007}, and bounding
the region by areas of high recombination rate, inferred using data
from Phase I, release 16a of the HapMap study \cite{mcvean-2004}, and
visualized in the UCSC Genome Browser
\cite{ucsc-genome-browser-2012}. All SNP information and genomic
positions are based on human genome assembly 17 (NCBI build 35).
\end{flushleft}
\label{table:associations} 
\end{table*}

Now we examine how the pathway enrichment findings can help us
identify additional genetic associations. The intuition is that,
having established that variants near genes in a pathway are more
likely to be associated with the phenotype, it is reasonable to
up-weight, or {\em prioritize,} these SNPs in the statistical
analysis. Our model-based framework achieves this by estimating an
enrichment parameter representing the increased prior probability that
SNPs assigned to the pathway are associated with the phenotype.  This
prior probability in turn raises the posterior probability of
association for these SNPs. This ability to reassess the evidence for
association of individual loci in light of enriched pathways is an
important feature of our model-based approach to pathway analysis, as
identifying individual disease-susceptibility loci is potentially more
informative than observing that a pathway is enriched for disease
associations. This is particularly the case for broad pathways, such
as cytokine signaling in immune system, which contains 225 genes, as
only a small proportion of these genes may actually harbour genetic
variants that affect Crohn's disease risk.

In simultaneous analysis of genetic variants based on Bayesian
variable selection, it is preferable, at least initially, to assess
evidence for associations across genomic regions, rather than for
individual SNPs. This is because when SNPs are correlated with one
another the association signal can be spread across these SNPs,
diluting the signal at any given SNP \cite{guan-stephens-2011}.
Therefore, we divide the genome into overlapping segments of 50 SNPs,
with an overlap of 25 SNPs between neighbouring segments. For each
segment, we compute $P_1$, the posterior probability that at least one
SNP in the segment is included in the multi-marker disease model. We
use segments with an equal number of SNPs so that, under the null
hypothesis of no enrichment, the prior probability that at least one
SNP is included is the same for every segment. Since a segment spans,
on average, 307 kb of the genome (98\% of segments are between 100 kb
and 1 Mb long), calculating $P_1$ for these segments provides only a
low-resolution map of genetic risk factors for Crohn's disease. Still,
this resolution suffices for the objectives of this case study. In
other applications, one could increase the resolution by calculating
PIPs for individual SNPs within selected regions. Hereafter, we use
$P_n$ to denote the posterior probability that at least $n$ SNPs are
included.

First, we compare Crohn's disease associations identified under the
null hypothesis that no pathways are enriched with associations
identified under the model in which a single pathway, cytokine
signaling in immune system, is enriched. Figure
\ref{fig:scatterplots}A shows how the evidence for association in each
segment ($P_1$) changes once we account for enrichment of cytokine
signaling genes. As expected, many segments (corresponding to points
above the diagonal in the scatterplot) show increased evidence for
association under the model in which cytokine signaling is
enriched. These segments contain SNPs assigned to the enriched
pathway. Segments corresponding to points on the diagonal show no
change in support for association, and these are segments that do not
contain SNPs assigned to the enriched pathway. Although it is not
clear from the figure due to over-plotting, most segments lie near the
bottom-left corner; when cytokine signaling is enriched, 17,261 out of
17,668 segments across the genome (97.7\%) have $P_1 \leq 0.1$.

Points in the top-right corner of Fig.~\ref{fig:scatterplots}A
correspond to regions with strong evidence for association even
without the benefit of feedback from pathway enrichment.  Genes
{\sf\em IL23R}, {\sf\em PTGER4}, {\sf\em ZNF365}, {\sf\em NKX2-3} and
{\sf\em ATG16L1} are not involved in cytokine signaling, nor are any
nearby genes, so the PIPs of SNPs near these genes are unaffected by
the hypothesis that cytokine signaling is enriched. {\sf\em NOD2}
(also known as {\sf\em CARD15}) and {\sf\em PTPN2} are cytokine
signaling genes, so these associations contribute to the evidence for
enrichment of this pathway, but because they already show strong
support for association without enrichment ($P_1$ is close to 1 under
the null hypothesis), these associations are not greatly affected by
enrichment. Reassuringly, these results recapitulate the strongest
associations reported in the original study (Table~3 in \cite{wtccc})
with trend {\em p}-values less than $4 \times 10^{-8}$, or those with
additive BFs greater than $10^{5.4}$. These results have also been
replicated in a follow-up study \cite{parkes-2007}, and have been
confirmed in meta-analyses with large combined sample sizes
\cite{barrett-2008, franke-2010}. Two additional regions at 3p21 and
5q33 are reported as associations in the original study \cite{wtccc},
although with trend-test {\em p}-values exceeding $5 \times
10^{-8}$. These Crohn's disease associations are replicated elsewhere
\cite{barrett-2008, franke-2010, parkes-2007}, but show only modest
evidence for association in our analysis; these regions are not
annotated to the cytokine signaling pathway, and the largest $P_1$ for
segments at these regions are $0.19$ and $0.21$, respectively.

Points near the top-left corner of Fig.~\ref{fig:scatterplots}A
correspond to regions of the genome that show strong support for
association only after accounting for enrichment of cytokine
signaling. Three additional regions stand out: the MHC class II
region, previously identified as a region showing moderate evidence of
association \cite{wtccc}; the {\sf\em IBD5} locus at 5q31, which
contains several candidate genes; and a region at position 17q21 near
gene {\sf\em STAT3}. Other genome-wide studies and meta-analyses
independently support Crohn's disease associations at these loci
\cite{barrett-2008, fernando-2008, franke-2010, mathew-2008,
  rioux-2000, silverberg-2007, vanheel-2004}. (See Supplementary
Results for further details on these loci.) In addition, two loci at
16q24 and 10q22 near genes {\sf\em IRF8} and {\sf\em CAMK2G} show
moderate support for association under the enrichment hypothesis;
$P_1$ is $0.49$ and $0.46$, respectively. Neither of these loci have
been identified as being significant associations in other Crohn's
disease studies. However, the association at locus 16q24, 84.45--84.6
Mb near {\sf\em IRF8} is potentially interesting, as it has
previously been identified in genome-wide studies of two other
autoimmune diseases, multiple sclerosis \cite{dejager-2009} and
systemic sclerosis \cite{gorlova-2011}. This gene belongs to a family
of transcription factors that regulate responses to type I interferons
(IFN-$\alpha$ and IFN-$\beta$), and these interferons are known play
critical roles in modulating inflammatory and immune responses to
pathogens \cite{oshea-2002}.

Next we investigate whether allowing for two enriched pathways reveals
any further associations. (See Methods for how $P_1$ is computed by
averaging over different models with two enriched pathways.) Figure
\ref{fig:scatterplots}B shows that allowing for enrichment of two
pathways does not yield compelling support for genetic associations
beyond those revealed by enrichment of cytokine signaling. However,
the segment with the greatest increase in $P_1$, from 0.04 to 0.53, is
at 158.4--159.1 Mb on chromosome 5, near gene {\sf\em IL12B}. This
locus was reported as a Crohn's disease association in
\cite{barrett-2008}, and was later confirmed in
\cite{franke-2010}. {\sf\em IL12B} is also associated with other
autoimmune diseases, including ulcerative colitis \cite{fisher-2008,
  franke-2008} and psoriasis \cite{cargill-2007}. Note that all
pathways listed in Table~\ref{table:ranking2} except cytokine
signaling implicate {\sf\em IL12B}, so the association signal at this
locus presumably contributes to evidence for enrichment of these
pathways.

Table~\ref{table:associations} summarizes the Crohn's disease
associations that are strongly supported by our analysis. Of the 10
regions listed in this table, 3 are revealed only after prioritizing
SNPs in enriched pathways. Each row in the table shows the SNP within
the critical region that has the largest PIP. In most cases, since
only a portion of all commonly occurring SNPs are included in the
study, this SNP is most likely in linkage disequilibrium with the
causal variant rather than being causal itself. We also show in this
table, for each selected region, evidence for multiple independent
risk factors, indicated by $P_2$, the posterior probability that at
least two SNPs in the region are independently associated with Crohn's
disease. Our calculations suggest the strong possibility of multiple
independent risk factors at the {\sf\em NOD2} locus, with $P_2 =
0.77$, a prediction that coincides with a previous study
\cite{hugot-2001}.

The large number of points approaching the middle of the $y$-axis in
Fig.~\ref{fig:scatterplots}A suggests that many other gene variants
involved in cytokine signaling may contribute to Crohn's disease
risk. In fact, the estimates $\bar{\theta}_0 = -4.11$ and
$\bar{\theta} = 1.96$ given in Table~\ref{table:ranking} imply that
roughly 1 out of every 140 SNPs in the pathway are expected to be
independent associations, for a total of about 47 independent risk
factors for Crohn's disease hidden among ``cytokine signaling in
immune system" genes. Our analysis has only identified a few of these
genes, which suggests that many more associations in this pathway
remain to be discovered. This prediction coincides to some extent with
a recent meta-analysis of Crohn's disease association studies
\cite{franke-2010}, as an additional 7 cytokine signaling
genes---{\sf\em IL1R1}, {\sf\em IP6K2}, {\sf\em JAK2}, {\sf\em IL2RA},
{\sf\em TYK2}, {\sf\em MAPK1} and {\sf\em MAP3K71P1}---overlap with
the susceptibility loci identified in this meta-analysis.

\subsection*{Sensitivity of pathway ranking to prior distribution 
of odds ratios}

Our approach requires specification of a prior distribution for the
odds ratios (see Methods). We assume a prior in which the log-odds
ratios ({\em i.e.} the coefficients $\beta_j$ in the additive model of
disease risk) are normally distributed with mean zero and standard
deviation $\sigma_a = 0.1$. This choice is based on odds ratios
reported in published genome-wide association studies (see
Methods). One concern is that slightly smaller or slightly larger
settings of $\sigma_a$ could also be justified, and these choices
could produce different results. Associations are unlikely to
accumulate at a greater rate in pathways that are not related to the
disease, even associations with small effects on disease risk, so we
predict that the ranking of pathway enrichments is largely robust to
the choice of $\sigma_a$.  Here we verify this claim. We assess the
sensitivity of our results to $\sigma_a$ by recomputing the BFs for
all candidate pathways with prior choices that favor slightly smaller
($\sigma_a = 0.06, 0.08$) and slightly larger coefficients ($\sigma_a
= 0.15, 0.2$).

\begin{figure}[t]
\centering
\includegraphics[width=2.75in,keepaspectratio=true]{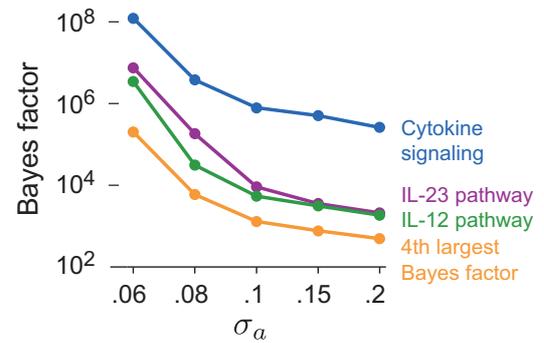}
\caption{{\bf The top four BFs for each setting of $\bm{\sigma_a}$.}
  In each case, the three largest BFs correspond, in order, to
  cytokine signaling in immune system, IL23-mediated signaling events,
  and IL12-mediated signaling events (these are also the top three
  pathways in Table~\ref{table:ranking}). The pathway with the
  fourth-largest BF differs across settings of $\sigma_a$.}
\label{fig:sensitivity-of-ranking} 
\end{figure}

Fig.~\ref{fig:sensitivity-of-ranking} shows that smaller settings of
$\sigma_a$ yield substantially more support for enrichment of
disease-related pathways, as expected. But the pathways with the
largest BFs are IL-23, IL-12 and cytokine signaling regardless of the
choice of $\sigma_a$. In the Supplementary Results, we show that the
BFs for most of the other 3158 candidate pathways do not change
substantially at different settings of $\sigma_a$. In summary, we
conclude that the pathway-level findings in this Crohn's disease
study are largely robust to priors that are not substantially
different from $\sigma_a = 0.1$.

\section*{Discussion}

Pathway analysis for genome-wide association studies has been as
advertised as a way to overcome some of the limitations of
conventional approaches to identifying genetic factors underlying
polygenic traits. Motivated by the observation that it is easier, in
principle, to identify associations within an enriched pathway, we
developed a model-based approach for simultaneously estimating
enrichment and prioritizing variants in enriched pathways. We
investigated the merits and limitations of this approach in a GWAS for
Crohn's disease. In this case study, we interrogated over 3,000
candidate pathways from several pathway databases, and confirmed the
importance of the IL-12 and IL-23 pathways in Crohn's disease
pathogenesis and, more broadly, the role of cytokines that mediate
immune responses. By prioritizing variants within the enriched
pathways identified in our analysis, we established strong support for
disease susceptibility loci beyond what are revealed by a conventional
analysis that ignores this pathway information. This suggests that
applying our methods to larger samples may reveal genetic loci that
have not yet been identified as risk factors for Crohn's disease and
other common diseases. Moreover, leveraging knowledge about variants
that modulate expression of genes (eQTLs) may also lead to stronger
evidence for pathway enrichments, and for associations within these
pathways.

Our approach to modeling enrichment was built on large-scale sparse
regression methods that have been applied to other problems in
statistics and genetics. The key idea behind our approach was to
introduce a parameter, $\theta$, that quantifies
enrichment---precisely, the increase in the probability that a SNP
assigned to a pathway is included in a polygenic model of the
phenotype. Given the generality of this approach, our method could be
useful for problems outside genetic association studies. One caveat is
that some of the approximations we used---approximations which help
scale the computations to hundreds of thousands of SNPs and thousands
of candidate pathways (see Supplementary Methods)---may not be
appropriate in some settings.

One attractive feature of our approach, which we illustrated in the
Crohn's disease case study, is that it can be used to assess how well
the data support enrichment of different combinations of multiple
pathways. Examining combinations of pathways for enrichment may
highlight interesting pathways that would not otherwise be highly
ranked. An example of this was the ``IL-12 and Stat4 dependent
signaling'' pathway, which showed little support for enrichment on its
own ($\BF = 18$), but became more interesting when considered jointly
for enrichment with cytokine signaling genes ($\BF = 2.4 \times
10^8$).

In contrast to many pathway analysis methods, we modeled pathway
enrichment at the level of variants, rather than genes. While there
are arguments for both approaches, a feature of the variant-based
approach is that, when there are multiple independent association
signals near a gene, all these signals contribute to the evidence for
enrichment of pathways containing this gene.

The Crohn's disease study was aimed at identifying common variants
associated with disease, but there is growing interest in using exome
and whole-genome sequencing to investigate the contribution of rare
variation to complex diseases and traits \cite{cirulli-2010,
  1000genomes, trynka-2011}. The computational complexity of our
method grows linearly with the number of SNPs, so it should scale well
to the large amount of data generated by high-throughput
sequencing. Since detecting associations with individual rare variants
is a hard problem \cite{bansal-2010, neale-2011}, pathway-based
analysis approaches such as ours that aggregate association signals
across sets of genes may play an important role in analysis of these
studies.

Currently, a drawback to our approach is that the prior variance of
additive effects on disease risk must be chosen beforehand. We based
our choice on the distribution of odds ratios reported in published
genome-wide association studies, and checked that the rankings of
enriched pathways were robust to different prior choices. Ideally, we
would estimate this parameter from the data instead, but, as we
discussed in Methods, we found that this did not work well for the
Crohn's disease data. In our estimation, the root of the problem is
that the log-odds ratios are not normally distributed (currently, we
assume that non-zero effects follow a normal distribution). One
possible solution would be to use a more flexible distribution for the
effect sizes, such as a mixture of two or more normals, but we have
not investigated this direction as it raises a number of questions
regarding modeling and computation. Despite this issue, we believe
that we have presented a useful framework for identifying enriched
pathways and genetic associations underlying these pathways.

\section*{Method}

\subsection*{Samples}

Our analysis is based on genome-wide marker data from a case-control
study with 1748 subjects affected by Crohn's disease, and 2938 control
subjects. The controls come from two groups: 1480 individuals from the
1958 Birth Cohort, and 1458 individuals from the UK Blood Services
cohort. All subjects are from Great Britain, and of self-described
European descent. Genetic associations from this study were first
reported in \cite{wtccc}.

All study subjects were genotyped for ${\about}500,000$ SNPs using a
commercial version of the Affymetrix GeneChip 500K platform.  We apply
quality control filters as described in \cite{wtccc}, and remove SNPs
that exhibit no variation in the sample. We discarded two additional
SNPs, rs1914328 on chromosome 8 and rs6601764 on chromosome 10, because
they show some evidence for association (single-SNP BFs
\cite{servin-stephens-2007} of $10^{3.5}$ and $10^{3.6}$,
respectively), but we cannot rule out the possibility of genotyping
errors as these associations are not supported by nearby SNPs (none of
the nearby SNPs have single-SNP BFs greater than 40). After
removing these two SNPs, we end up with 442,001 SNPs on autosomal
chromosomes. We estimate missing genotypes at these SNPs using the
mean posterior minor allele count from BIMBAM \cite{guan-stephens-2008,
  servin-stephens-2007}, with SNP data from Phase II of the
International HapMap Consortium project \cite{hapmap}. To be
consistent with the original analysis, refSNP identifiers and
locations of SNPs are based on human genome reference assembly 17
(NCBI build 35).

\subsection*{Population stratification}

Analysis of pathways should, in principle, be robust to population
stratification because spurious associations that arise from
population structure are unlikely to accumulate at a greater rate in
the pathway---recall, we define the enrichment parameter as the
increase in proportion of variants associated with disease risk
relative to the proportion genome-wide. However, population
stratification could still produce individual false positive
associations, either inside or outside enriched pathways, so in
general one should account for this in the analysis
\cite{clayton-2005, mccarthy-2008, pearson-manolio-2008, price-2010,
  tintle-2011}. For the Crohn's disease study, the original report
\cite{wtccc} and subsequent analyses \cite{barrett-2008, zeggini-2008}
affirm that cryptic population structure does not have a substantive
impact on the analysis. Thus we do not attempt to correct for population
structure in our analysis.

\subsection*{Pathways, and assignment of SNPs to pathways}
\label{sec:pathways}

\begin{table*}[t]
\begin{center}
\caption{\bf Overview of pathways used in the analysis}
\label{table:pathways} 
{\sf\begin{tabular}{m{1.7in}clr}
database & refs. & download location & \#pathways \\ \hline
\multirow{2}{*}{BioCarta} & & 
\url{www.openbioinfor-} & \multirow{2}{*}{298} \\[-0.5ex]
& & \url{matics.org/gengen} \\
\shaderow
Cancer Cell Map & & Pathway Commons & 10 \\
\multirow{2}{*}{HumanCyc} & 
\multirow{2}{*}{\citenum{biocyc-2010,humancyc-2004}}
& NCBI BioSystems & 54 \\
& & Pathway Commons & 224 \\
\shaderow
\raggedright Kyoto Encyclopedia of Genes and Genomes (KEGG) & 
\citenum{kegg-2010} & NCBI BioSystems & 399 \\
NCI Nature Pathway & \multirow{2}{*}{\citenum{pid-2009}} & 
NCBI BioSystems & 186  \\
Interaction Database (PID) & & Pathway Commons & 179  \\
\shaderow
PANTHER & \citenum{panther-2009,panther-2010} & 
\url{www.pantherdb.org} & 128 \\
\multirow{2}{*}{Reactome} & \multirow{2}{*}{\citenum{reactome-2011}} & 
NCBI BioSystems & 1093 \\
& & Pathway Commons & 1070 \\
\shaderow
WikiPathways & \citenum{wikipathways-2011,wikipathways-2008} & 
NCBI BioSystems & 147
\end{tabular}}
\\[1em]
From these 3788 pathways, we obtain 3158 unique gene sets.
\end{center}
\end{table*}

We aim for a comprehensive evaluation of pathways accessible on the
Web in standard, computer-readable formats
\cite{pathguide-2006, bauer-mehren-2009, biopax-2010}. Since the results
hinge on the quality of the pathways used in our analysis, we restrict
the analysis to curated, peer-reviewed pathways based on experimental
evidence, and pathways inferred via gene homology. We draw candidate
pathways from the collections listed in Table~\ref{table:pathways}
(see Supplementary Methods for details). KEGG and HumanCyc are
primarily databases of metabolic pathways, and are unlikely to be
relevant to Crohn's disease, but we include them for completeness.

Since we combine pathways from different sources, we encounter
pathways with inconsistent definitions \cite{soh-2010,stobbe-2011}.
There is no single explanation for the lack of consensus in pathway
definitions, and we have no reason to prefer one definition over
another, so we include multiple versions of a pathway in our
analysis. Many of the pathways in these databases are arranged
hierarchically; we incorporate all elements of the hierarchy into our
analysis. We treat each candidate pathway as a set of genes, ignoring
details such as molecules involved in biochemical reactions, and
cellular locations of these reactions. From 3788 total pathways
(Table~\ref{table:pathways}), we obtain 3158 unique gene sets. With
these pathways we achieve ${\about}39\%$ gene coverage (see
Supplementary Results).

Based on findings that the majority of variants modulating gene
expression lie within 100 kb of the gene's transcribed region
\cite{cookson-2009, dixon-2007, stranger-2007}, we assign a SNP to a
gene if it is within 100 kb of the transcribed region. Others have
opted for a 20 kb window \cite{holmans-2009, wang-2009} based on
findings that {\em cis}-acting expression QTLs are rarely more than 20
kb from the gene \cite{veyrieras-2008}. We prefer a more inclusive
mapping of SNPs to genes, since the benefit of including potentially
relevant SNPs in a pathway when the association signal is sparse seems
likely to outweigh the cost of including a larger number of irrelevant
markers.

\subsection*{Statistical analysis}

The Bayesian variable selection approach to simultaneous interrogation
of SNPs involves fitting the multi-marker disease model to the data
using different combinations of SNPs. Fitting all markers
simultaneously weeds out multiple associations at markers that are in
linkage disequilibrium with one another, leaving only one marker for
each {\em independent association}---an association that signals a
variant contributing to disease risk independently of other
risk-conferring variants.

\subsubsection*{Likelihood} 

The likelihood specifies the probability of observing the phenotype
observations $y$ given the genotypes $\mathbf{X}$, the intercept
$\beta_0$, and the regression coefficients $\beta = (\beta_1, \ldots,
\beta_p)$. From the additive model for the log-odds of disease
(eq.~\ref{eq:logistic}), $p_i = \psi\big(\beta_0 + \sum_{j=1}^p
x_{ij}\beta_j \big)$ is the probability that $y_i = 1$, in which $\psi(x)
= 1/(1+e^{-x})$ is the sigmoid function. Assuming independence of the
observations $y_i$, the likelihood is
\begin{align}
p(y \,|\, \mathbf{X}, \beta_0, \beta) = 
\prod_{i=1}^n p_i^{y_i} (1-p_i)^{1-y_i}.
\label{eq:likelihood-logistic}
\end{align}

\subsubsection*{Priors}
\label{sec:priors}
 
Next we specify prior distributions for the genome-wide log-odds
$\theta_0$, the enrichment parameter $\theta$, the intercept
$\beta_0$, and the coefficients $\beta_j$ of SNPs included in the
multi-marker model of disease risk.

Since inferences strongly depend on $\theta_0$, and since $\theta_0$
is unknown and will be different for each setting, we estimate this
parameter from the data rather than fix it {\em a priori}. Following
\cite{bvs, guan-stephens-2011}, we assign a uniform prior to
$\theta_0$. We restrict $\theta_0$ to $[-6,-2]$, so as few as 0 and as
many as ${\about}4400$ SNPs are expected to be included {\em a
  priori.}

We place a uniform prior on $\theta$, restricted to $[0,3]$. This
prior permits a wide range of enrichments because, in our view,
enrichments greater than a thousand-fold are unlikely to occur. Note
that we do not allow negative enrichments; that is, we do not consider
pathways that are underrepresented for associations with the
phenotype. Negative enrichments could potentially be useful in other
scenarios, but for most genome-wide association studies where there
are generally few significant associations to begin with, negative
pathway enrichments are difficult to find and are unlikely to have a
useful interpretation.

For the prior on the non-zero coefficients $\beta_j$, we follow a
standard practice that assumes they are {\em i.i.d.} normal with zero
mean and standard deviation $\sigma_a$
\cite{george-mcculloch-1993,mitchell-beauchamp-1988}. Ordinarily, to
combat sensitivity of the results to the choice of $\sigma_a$, we
would place a prior on $\sigma_a$ and integrate over this parameter to
let the data drive selection of $\sigma_a$. This approach is taken in
\cite{bvs,guan-stephens-2011}. But in our case, we find that the
heterogeneity of the odds ratios in Crohn's disease presents a
problem: although we expect most odds ratios for a common
disease---and specifically odds ratios in a pathway relevant to
disease pathogenesis---to be close to 1, the odds ratios corresponding
to the strongest Crohn's disease associations drive estimates of
$\sigma_a$ toward larger values, and a normal distribution that puts
too little weight on modest odds ratios. One possible strategy would
be to redo the analysis after removing associated regions with the
largest odds ratios, but this is an unattractive solution because SNPs
with large odds ratios would not contribute to the evidence for
enrichment. Instead, we fix $\sigma_a$, grounding the choice on
typical odds ratios reported in published genome-wide association
studies, and we assess the robustness of our findings to this choice.
(There is the potential for more principled solutions; see
Discussion.) Our choice is $\sigma_a = 0.1$, which favours odds ratios
close to 1 (95\% of the odds ratios lie between 0.82 and 1.22 {\em a
  priori}), while being large enough to capture a significant fraction
of the odds ratios for common alleles reported in genome-wide
association studies of complex disease traits. According to a recent
review \cite{bodmer-bonilla-2008}, approximately 40\% of estimated
odds ratios are between 1.1 and 1.2, and an additional $10\%$ of odds
ratios are smaller than 1.1. This prior also closely corresponds to a
survey of odds ratios reported in genetic association studies of
common diseases \cite{ioannidis-2006}. Since there may be
justification for a slightly smaller or slightly larger $\sigma_a$, we
also try different values for $\sigma_a$, and examine how these
choices affect the ranking of enriched pathways (see Results).

To complete the probability model, we assign an improper uniform prior
to the intercept, $\beta_0$. In general, one must be careful with use
of improper priors in Bayesian variable selection because they can
result in improper posteriors. A sufficient condition for a proper
posterior, and a well-defined BF, with logistic regression
(eq.~\ref{eq:logistic}) is that the maximum likelihood estimator of
$\beta$ conditioned on which variables are included in the model, and
on the other model parameters, is unique and finite
\cite{obrien-2004}.  Unfortunately, this condition is difficult to
check exhaustively. But we can at least guarantee that the posterior
is proper under the variational approximation (see Supplementary
Methods) so long as the coordinate ascent steps converge to a unique
solution.

\subsubsection*{Bayes factors and posterior odds}

Ideally, we would assess evidence for an enrichment model by computing
the posterior probability for that model. But computing these
posterior probabilities is impractical for several reasons, one being
that it would involve computations for a large number of combinations
of pathways. Instead, we compute a Bayes factor \eqref{eq:BF} for each
candidate pathway, or combination of pathways, then we weigh the Bayes
factor against the prior odds to obtain the posterior odds for the
enrichment model:
\begin{align}
\overbracket[0pt]{\frac{p(\theta > 0 \,|\, \mathbf{X}, y, a)}
{p(\theta = 0 \,|\, \mathbf{X}, y, a)}}^{\sf (posterior\,odds)} =
\overbracket[0pt]{\frac{p(y \,|\, \mathbf{X}, a, \theta > 0)}
{p(y \,|\, \mathbf{X}, a, \theta = 0)}}^{\sf (Bayes\,factor)}
 \times 
\overbracket[0pt]{\frac{p(\theta > 0)}{p(\theta = 0)}}^{\sf (prior\,odds)}.
\label{eq:posterior-odds}
\end{align}

In our pathway enrichment results (Tables~\ref{table:ranking} and
\ref{table:ranking2}), we report BFs rather than posterior odds.
Although our results would be more straightforward to interpret had we
provided posterior odds instead of BFs, posterior odds are easily
obtained from BFs, and reporting BFs offers the reader flexibility to
judge the evidence based on his or her own prior. In the results, we
illustrate how to gauge support for each enrichment hypothesis by
weighing against a ``conservative'' prior. However, the reader may
have reason to choose a different prior, such as one that favours
certain pathways above others.

To allow for uncertainty in $\theta_0$ and $\theta$ when evaluating the
BFs, the likelihood under the enrichment hypothesis ($\theta > 0$) and
the likelihood under the null ($\theta = 0$) are each expressed as an
average over possible assignments to $\theta_0$ and $\theta$:
\begin{align}
\BF(a) = \frac{\iint p(y \,|\, \mathbf{X}, a, \theta_0, \theta) \, 
            p(\theta_0) \, p(\theta) \, d\theta \, d\theta_0}
           {\int p(y \,|\, \mathbf{X}, a, \theta_0, \theta = 0) \, 
            p(\theta_0) \, d\theta_0}.
\label{eq:BF-2}
\end{align}
Each instance of $p(y \,|\, \mathbf{X}, a, \theta_0, \theta)$ in
\eqref{eq:BF-2} expands as an average over possible assignments
to the intercept $\beta_0$ and regression coefficients $\beta$:
\begin{align}
p(y \,|\, \mathbf{X}, a, \theta_0, \theta) &= 
{\textstyle \iint p(y \,|\, \mathbf{X}, \beta_0, \beta)} \, p(\beta_0) \,
\nonumber \\ & \qquad
\times \prod_{j=1}^p 
p(\beta_j \,|\, a_j, \theta_0, \theta) \, d\beta_0 \, d\beta.
\label{eq:likelihood-theta}
\end{align}
Factor $p(\beta_j \,|\, a_j, \theta_0, \theta) = \pi_j N(0,\sigma_a^2)
+ (1-\pi_j) \, \delta_0$ is the ``spike and slab'' prior
\cite{george-mcculloch-1993,mitchell-beauchamp-1988}, in which $\pi_j
= p(\beta_j \neq 0)$ is determined according to
\eqref{eq:pathway-prior}. Here, $\delta_0$ denotes the delta mass, or
``spike'', at zero, $N(\mu,\sigma^2)$ is the normal density with mean
$\mu$ and variance $\sigma^2$, $p(\beta_0)$ is the (improper) uniform
prior, and $p(y \,|\,\mathbf{X}, \beta_0, \beta)$ is the logistic
regression likelihood \eqref{eq:likelihood-logistic}. Computation of
Bayes factors is described in Supplementary Methods.

\subsubsection*{Posterior inclusion probabilities and other statistics}

Here we define PIPs and other posterior quantities used in our
analysis for the case when a pathway, or combination of pathways, is
enriched. Posterior statistics under the null hypothesis are obtained
by setting $\theta = 0$.

Like the BFs, the PIPs are obtained by averaging over $\theta_0$ and
$\theta$. Taking $\Data = \{\mathbf{X}, y, a\}$ as shorthand for all
the data, we have
\begin{align}
\PIP(j) &\equiv p(\beta_j \neq 0 \,|\, \Data) \nonumber \\ &= 
\textstyle \iint p(\beta_j \neq 0 \,|\, \Data, \theta_0, \theta) \,
p(\theta_0, \theta \,|\, \Data) \, d\theta_0 \, d\theta.
\label{eq:PIP}
\end{align}

To identify regions of the genome associated with disease risk, we
calculate, for each region, the posterior probability that at least
one SNP in the region is included in the multi-marker disease model
(see Results for an explanation). Let $S = n$ represent the event
that exactly $n$ SNPs in a given region are included in the
multi-marker disease model, so that $P_1 = p(S \geq 1 \,|\,
\Data)$. These posterior statistics are easily calculated from the
PIPs using the variational approximation (see Supplementary Methods).

Since no pair of pathways stands out in Table~\ref{table:ranking2} as
having greater support than any other pair, we compute the posterior
statistics $P_1$ by averaging over the different models with two
enriched pathways, including models with BFs too small to be included
in the table, weighting these models by their BFs. Implicitly, this
assumes that all models with two enriched pathways are equally
plausible {\em a priori.} The ability to average across models in this
way is an advantage of adopting the Bayesian approach to model
comparison, because it allows us to assess genetic associations in
light of the enrichment evidence without having to choose a single
pair of pathways.  Suppose we have $m$ enrichment models $a^{(1)},
\ldots, a^{(m)}$ with corresponding Bayes factors $\BF(a^{(1)}),
\ldots, \BF(a^{(m)})$, then $P_1$ for a given region is
\begin{align}
P_1 = 
\frac{\sum_{i=1}^m p(S \geq 1 \,|\, \Data, a^{(i)}) \, \BF(a^{(i)})}
     {\sum_{i=1}^m \BF(a^{(i)})}. 
\label{eq:P1-two-pathways}
\end{align}
Further details about computation of posterior statistics are given
in Supplementary Methods.

\section*{Acknowledgments}

Thanks to Yongtao Guan for assistance with the Crohn's disease case
study, Kevin Bullaughey and John Zekos for expert technical support,
Emek Demir, Lewis Geer, Benjamin Cross and the rest of the Pathway
Commons team for help with pathway databases, and Gorka
Alkorta-Aranburu, Niall Cardin, Anna Di Rienzo, Hariklia
Eleftherohorinou, Timoth\'{e}e Flutre, Stoyan Georgiev, Ron Hause,
Bryan Howie, Ellen Leffler, Dan Nicolae, Heejung Shim, Xiaoquan Wen
and Xiang Zhou for helpful discussions. This work was supported by a
grant from the National Institute of Health (HG02585), and a
cross-disciplinary postdoctoral fellowship from the Human Frontiers
Science Program.


\bibliography{pathway}

\begin{thebibliography}{100}

\bibitem{altshuler-2008}
D.~Altshuler, M.~J. Daly, and E.~S. Lander.
\newblock Genetic mapping in human disease.
\newblock {\em Science}, 322(5903):881--888, 2008.

\bibitem{frazer-2009}
K.~A. Frazer, S.~S. Murray, N.~J. Schork, and E.~J. Topol.
\newblock Human genetic variation and its contribution to complex traits.
\newblock {\em Nature Reviews Genetics}, 10(4):241--251, 2009.

\bibitem{mccarthy-2008}
M.~I. McCarthy, G.~R. Abecasis, L.~R. Cardon, D.~B. Goldstein, J.~Little,
  J.~P.~A. Ioannidis, and J.~N. Hirschhorn.
\newblock Genome-wide association studies for complex traits: consensus,
  uncertainty and challenges.
\newblock {\em Nature Reviews Genetics}, 9(5):356--369, 2008.

\bibitem{pearson-manolio-2008}
T.~A. Pearson and T.~A. Manolio.
\newblock How to interpret a genome-wide association study.
\newblock {\em Journal of the American Medical Association},
  299(11):1335--1344, 2008.

\bibitem{abraham-2009c}
C.~Abraham and J.~H. Cho.
\newblock Inflammatory bowel disease.
\newblock {\em New England Journal of Medicine}, 361(21):2066--2078, 2009.

\bibitem{barrett-2008}
J.~C. Barrett, S.~Hansoul, D.~L. Nicolae, J.~H. Cho, et~al.
\newblock Genome-wide association defines more than 30 distinct susceptibility
  loci for {Crohn's} disease.
\newblock {\em Nature Genetics}, 40:955--962, 2008.

\bibitem{franke-2010}
A.~Franke, D.~P.~B. McGovern, J.~C. Barrett, K.~Wang, et~al.
\newblock Genome-wide meta-analysis increases to 71 the number of confirmed
  {Crohn's} disease susceptibility loci.
\newblock {\em Nature Genetics}, 42(12):1118--1125, 2010.

\bibitem{khor-2011}
B.~Khor, A.~Gardet, and R.~J. Xavier.
\newblock Genetics and pathogenesis of inflammatory bowel disease.
\newblock {\em Nature}, 474(7351):307--317, 2011.

\bibitem{stappenbeck-2011}
T.~S. Stappenbeck, J.~D. Rioux, A.~Mizoguchi, T.~Saitoh, A.~Huett,
  A.~Darfeuille-Michaud, T.~Wileman, N.~Mizushima, S.~Carding, S.~Akira,
  M.~Parkes, and R.~J. Xavier.
\newblock Crohn disease: a current perspective on genetics, autophagy and
  immunity.
\newblock {\em Autophagy}, 7(4):355--374, 2011.

\bibitem{vanlimbergen-2009}
J.~{Van Limbergen}, D.~C. Wilson, and J.~Satsangi.
\newblock The genetics of {Crohn's} disease.
\newblock {\em Annual Review of Genomics and Human Genetics}, 10(1):89--116,
  2009.

\bibitem{eichler-2010}
E.~E. Eichler, J.~Flint, G.~Gibson, A.~Kong, S.~M. Leal, J.~H. Moore, and J.~H.
  Nadeau.
\newblock Missing heritability and strategies for finding the underlying causes
  of complex disease.
\newblock {\em Nature Reviews Genetics}, 11(6):446--450, 2010.

\bibitem{manolio-2009}
T.~A. Manolio, F.~S. Collins, N.~J. Cox, D.~B. Goldstein, et~al.
\newblock Finding the missing heritability of complex diseases.
\newblock {\em Nature}, 461:747--753, 2009.

\bibitem{rioux-abbas-2005}
J.~D. Rioux and A.~K. Abbas.
\newblock Paths to understanding the genetic basis of autoimmune disease.
\newblock {\em Nature}, 435(7042):584--9, 2005.

\bibitem{ropers-2007}
H-H. Ropers.
\newblock New perspectives for the elucidation of genetic disorders.
\newblock {\em American Journal of Human Genetics}, 81:199--207, 2007.

\bibitem{bodmer-bonilla-2008}
W.~Bodmer and C.~Bonilla.
\newblock Common and rare variants in multifactorial susceptibility to common
  diseases.
\newblock {\em Nature Genetics}, 40(6):695--701, 2008.

\bibitem{pritchard-2001}
J.~K. Pritchard.
\newblock Are rare variants responsible for susceptibility to complex diseases?
\newblock {\em American Journal of Human Genetics}, 69(1):124--137, 2001.

\bibitem{ballard-2010}
D.~Ballard, C.~Abraham, J.~Cho, and H.~Zhao.
\newblock Pathway analysis comparison using {Crohn's} disease genome wide
  association studies.
\newblock {\em BMC Medical Genomics}, 3(1):25, 2010.

\bibitem{braun-2011}
R.~Braun and K.~Buetow.
\newblock Pathways of distinction analysis: a new technique for multi-{SNP}
  analysis of {GWAS} data.
\newblock {\em PLoS Genetics}, 7(6):e1002101, 2011.

\bibitem{ramanan-2012}
V.~K. Ramanan, L.~Shen, J.~H. Moore, and A.~J. Saykin.
\newblock Pathway analysis of genomic data: concepts, methods and prospects for
  future development.
\newblock {\em Trends in Genetics}, pages 1--10, 2012.

\bibitem{wang-2010}
K.~Wang, M.~Li, and H.~Hakonarson.
\newblock Analysing biological pathways in genome-wide association studies.
\newblock {\em Nature Reviews Genetics}, 11(12):843--854, 2010.

\bibitem{yaspan-2011}
B.~L. Yaspan and O.~J. Veatch.
\newblock Strategies for pathway analysis from {GWAS} data.
\newblock {\em Current Protocols in Human Genetics}, 71:1.20.1--1.20.15, 2011.

\bibitem{cantor-2010}
R.~M. Cantor, K.~Lange, and J.~S. Sinsheimer.
\newblock Prioritizing {GWAS} results: a review of statistical methods and
  recommendations for their application.
\newblock {\em American Journal of Human Henetics}, 86:6--22, 2010.

\bibitem{hartwell-2004}
L.~Hartwell.
\newblock Robust interactions.
\newblock {\em Science}, 303(5659):774--775, 2004.

\bibitem{hirschhorn-2009}
J.~N. Hirschhorn.
\newblock Genomewide association studies---illuminating biologic pathways.
\newblock {\em New England Journal of Medicine}, 360:1699--1701, 2009.

\bibitem{schadt-2009}
E.~E. Schadt.
\newblock Molecular networks as sensors and drivers of common human diseases.
\newblock {\em Nature}, 461(7261):218--223, 2009.

\bibitem{aerts-2006}
S.~Aerts, D.~Lambrechts, S.~Maity, P.~{Van Loo}, B.~Coessens, F.~{De Smet},
  L.~Tranchevent, B.~{De Moor}, P.~Marynen, B.~Hassan, P.~Carmeliet, and
  Y.~Moreau.
\newblock Gene prioritization through genomic data fusion.
\newblock {\em Nature Biotechnology}, 24(5):537--544, 2006.

\bibitem{baranzini-2009}
S.~E. Baranzini, N.~W. Galwey, J.~Wang, P.~Khankhanian, et~al.
\newblock Pathway and network-based analysis of genome-wide association studies
  in multiple sclerosis.
\newblock {\em Human Molecular Genetics}, 18(11):2078--2090, 2009.

\bibitem{chen-2011}
M.~Chen, J.~Cho, and H.~Zhao.
\newblock Incorporating biological pathways via a {Markov} random field model
  in genome-wide association studies.
\newblock {\em PLoS Genetics}, 7(4):e1001353, 2011.

\bibitem{franke-2006}
L.~Franke, H.~van Bakel, L.~Fokkens, E.~D. de~Jong, M.~Egmont-Petersen, and
  C.~Wijmenga.
\newblock Reconstruction of a functional human gene network, with an
  application for prioritizing positional candidate genes.
\newblock {\em American Journal of Human Genetics}, 78(6):1011--1025, 2006.

\bibitem{lage-2007}
K.~Lage, E.~O. Karlberg, Z.~M. Storling, P.~I. Olason, A.~G. Pedersen,
  O.~Rigina, A.~M. Hinsby, Z.~Tumer, F.~Pociot, N.~Tommerup, Y.~Moreau, and
  S.~Brunak.
\newblock A human phenome-interactome network of protein complexes implicated
  in genetic disorders.
\newblock {\em Nature Biotechnology}, 25:309--316, 2007.

\bibitem{raychaudhuri-2009}
S.~Raychaudhuri, R.~M. Plenge, E.~J. Rossin, A.~C.~Y. Ng, S.~M. Purcell,
  P.~Sklar, E.~M. Scolnick, R.~J. Xavier, D.~Altshuler, and M.~J. Daly.
\newblock Identifying relationships among genomic disease regions: predicting
  genes at pathogenic {SNP} associations and rare deletions.
\newblock {\em PLoS Genetics}, 5(6):e1000534, 2009.

\bibitem{saccone-2008}
S.~F. Saccone, N.~L. Saccone, G.~E. Swan, P.~A.~F. Madden, A.~M. Goate, J.~P.
  Rice, and L.~J. Bierut.
\newblock Systematic biological prioritization after a genome-wide association
  study: an application to nicotine dependence.
\newblock {\em Bioinformatics}, 24(16):1805--1811, 2008.

\bibitem{tranchevent-2011}
L.~Tranchevent, F.~Bonachela Capdevila, D.~Nitsch, B.~{De Moor}, P.~{De
  Causmaecker}, and Y.~Moreau.
\newblock A guide to web tools to prioritize candidate genes.
\newblock {\em Briefings in Bioinformatics}, 12(1):22--32, 2011.

\bibitem{wu-2008}
X.~Wu, R.~Jiang, M.~Q. Zhang, and S.~Li.
\newblock Network-based global inference of human disease genes.
\newblock {\em Molecular Systems Biology}, 4:189, 2008.

\bibitem{bottolo-richardson-2010}
L.~Bottolo and S.~Richardson.
\newblock Evolutionary stochastic search for {Bayesian} model exploration.
\newblock {\em Bayesian Analysis}, 5:583--618, 2010.

\bibitem{bvs}
P.~Carbonetto and M.~Stephens.
\newblock Scalable variational inference for {Bayesian} variable selection in
  regression, and its accuracy in genetic association studies.
\newblock {\em Bayesian Analysis}, 7:73--108, 2012.

\bibitem{guan-stephens-2011}
Y.~Guan and M.~Stephens.
\newblock Bayesian variable selection regression for genome-wide association
  studies, and other large-scale problems.
\newblock {\em Annals of Applied Statistics}, 5(3):1780--1815, 2011.

\bibitem{he-lin-2011}
Q.~He and D.~Lin.
\newblock A variable selection method for genome-wide association studies.
\newblock {\em Bioinformatics}, 27(1):1--8, 2011.

\bibitem{hoggart-2008}
C.~J. Hoggart, J.~C. Whittaker, M.~De~Iorio, and D.~J. Balding.
\newblock Simultaneous analysis of all {SNPs} in genome-wide and re-sequencing
  association studies.
\newblock {\em PLoS Genetics}, 4(7):e1000130, 2008.

\bibitem{logsdon}
B.~A. Logsdon, G.~E. Hoffman, and J.~G. Mezey.
\newblock A variational {Bayes} algorithm for fast and accurate multiple locus
  genome-wide association analysis.
\newblock {\em BMC Bioinformatics}, 11(1):58, 2010.

\bibitem{segura-2012}
V.~Segura, B.~J. Vilhj{\'a}lmsson, A.~Platt, A.~Korte, {\"U}mit Seren, Q.~Long,
  and M.~Nordborg.
\newblock An efficient multi-locus mixed-model approach for genome-wide
  association studies in structured populations.
\newblock {\em Nature Genetics}, 44(7):825--830, 2012.

\bibitem{yi-2008}
N.~Yi and S.~Xu.
\newblock Bayesian {Lasso} for quantitative trait loci mapping.
\newblock {\em Genetics}, 179:1045–--1055, 2008.

\bibitem{wu-2009}
T.~T. Wu, Y.~F. Chen, T.~Hastie, E.~Sobel, and K.~Lange.
\newblock Genome-wide association analysis by {Lasso} penalized logistic
  regression.
\newblock {\em Bioinformatics}, 25(6):714--721, 2009.

\bibitem{platt-2010}
A.~Platt, B.~J Vilhj{\'a}lmsson, and M.~Nordborg.
\newblock Conditions under which genome-wide association studies will be
  positively misleading.
\newblock {\em Genetics}, 186(3):1045--1052, 2010.

\bibitem{gaffney-2012}
D.~J. Gaffney, J.~Veyrieras, J.~F. Degner, P.~Roger, A.~A. Pai, G.~E. Crawford,
  M.~Stephens, Y.~Gilad, and J.~K. Pritchard.
\newblock Dissecting the regulatory architecture of gene expression {QTLs}.
\newblock {\em Genome Biology}, 13(1):R7, 2012.

\bibitem{lee-2009}
S.~Lee, A.~M. Dudley, D.~Drubin, P.~A. Silver, N.~J. Krogan, D.~Pe'er, and
  D.~Koller.
\newblock Learning a prior on regulatory potential from {eQTL} data.
\newblock {\em PLoS Genetics}, 5(1), 2009.

\bibitem{lewinger-2007}
J.~P. Lewinger, D.~V. Conti, J.~W. Baurley, T.~J. Triche, and D.~C. Thomas.
\newblock Hierarchical {Bayes} prioritization of marker associations from a
  genome-wide association scan for further investigation.
\newblock {\em Genetic Epidemiology}, 31:871--883, 2007.

\bibitem{veyrieras-2008}
J-B. Veyrieras, S.~Kudaravalli, Su~Y. Kim, E.~T. Dermitzakis, Y.~Gilad,
  M.~Stephens, and J.~K. Pritchard.
\newblock High-resolution mapping of {Expression-QTLs} yields insight into
  human gene regulation.
\newblock {\em PLoS Genetics}, 4(10):e1000214, 2008.

\bibitem{pathguide-2006}
G.~D. Bader, M.~P. Cary, and C.~Sander.
\newblock Pathguide: a pathway resource list.
\newblock {\em Nucleic Acids Research}, 34(S1):D504--D506, 2006.

\bibitem{bauer-mehren-2009}
A.~Bauer-Mehren, L.~I. Furlong, and F.~Sanz.
\newblock Pathway databases and tools for their exploitation: benefits, current
  limitations and challenges.
\newblock {\em Molecular Systems Biology}, 5:290, 2009.

\bibitem{wtccc}
{Wellcome Trust Case Control Consortium}.
\newblock Genome-wide association study of {14,000} cases of seven common
  diseases and {3,000} shared controls.
\newblock {\em Nature}, 447:661--678, 2007.

\bibitem{holmans-2009}
P.~Holmans, E.~K. Green, J.~S. Pahwa, M.~A.~R. Ferreira, S.~M. Purcell,
  P.~Sklar, M.~J. Owen, M.~C. O'Donovan, and N.~Craddock.
\newblock {Gene Ontology} analysis of {GWA} study data sets provides insights
  into the biology of bipolar disorder.
\newblock {\em American Journal of Human Genetics}, 85(1):13--24, 2009.

\bibitem{peng-2010}
G.~Peng, L.~Luo, H.~Siu, Y.~Zhu, P.~Hu, S.~Hong, J.~Zhao, X.~Zhou, J.~D.
  Reveille, L.~Jin, C.~I. Amos, and M.~Xiong.
\newblock Gene and pathway-based second-wave analysis of genome-wide
  association studies.
\newblock {\em European Journal of Human Genetics}, 18(1):111--117, 2010.

\bibitem{torkamani-2008}
A.~Torkamani, E.~J. Topol, and N.~J. Schork.
\newblock Pathway analysis of seven common diseases assessed by genome-wide
  association.
\newblock {\em Genomics}, 92(5):265--272, 2008.

\bibitem{wang-2009}
K.~Wang, H.~Zhang, S.~Kugathasan, V.~Annese, J.~P. Bradfield, et~al.
\newblock Diverse genome-wide association studies associate the {IL12/IL23}
  pathway with {Crohn} disease.
\newblock {\em American Journal of Human Genetics}, 84(3):399--405, 2009.

\bibitem{abraham-cho-2009b}
C.~Abraham and J.~Cho.
\newblock {Interleukin-23/Th17} pathways and inflammatory bowel disease.
\newblock {\em Inflammatory Bowel Diseases}, 15(7):1090--1100, 2009.

\bibitem{li-2009}
Y.~Li, C.~Willer, S.~Sanna, and G.~Abecasis.
\newblock Genotype imputation.
\newblock {\em Annual Review of Genomics and Human Genetics}, 10:387--406,
  2009.

\bibitem{marchini-2010}
J.~Marchini and B.~Howie.
\newblock Genotype imputation for genome-wide association studies.
\newblock {\em Nature Reviews Genetics}, 11(7):499--511, 2010.

\bibitem{servin-stephens-2007}
B.~Servin and M.~Stephens.
\newblock Imputation-based analysis of association studies: candidate regions
  and quantitative traits.
\newblock {\em PLoS Genetics}, 3(7):e114, 2007.

\bibitem{kass-raftery-1995}
R.~E. Kass and A.~E. Raftery.
\newblock Bayes factors.
\newblock {\em Journal of the American Statistical Association}, 90:773--795,
  1995.

\bibitem{stephens-balding-2009}
M.~Stephens and D.~J. Balding.
\newblock Bayesian statistical methods for genetic association studies.
\newblock {\em Nature Reviews Genetics}, 10:681--690, 2009.

\bibitem{parkes-2007}
M.~Parkes, J.~C. Barrett, N.~J. Prescott, M.~Tremelling, et~al.
\newblock Sequence variants in the autophagy gene {IRGM} and multiple other
  replicating loci contribute to {Crohn's} disease susceptibility.
\newblock {\em Nature Genetics}, 39(7):830--832, 2007.

\bibitem{reactome-2011}
D.~Croft, G.~O'Kelly, G.~Wu, R.~Haw, et~al.
\newblock Reactome: a database of reactions, pathways and biological processes.
\newblock {\em Nucleic Acids Research}, 39(S1):D691--D697, 2011.

\bibitem{pid-2009}
C.~F. Schaefer, K.~Anthony, S.~Krupa, J.~Buchoff, M.~Day, T.~Hannay, and K.~H.
  Buetow.
\newblock {PID}: the {Pathway Interaction Database}.
\newblock {\em Nucleic Acids Research}, 37(S1):D674--D679, 2009.

\bibitem{biosystems-2010}
L.~Y. Geer, A.~Marchler-Bauer, R.~C. Geer, L.~Han, J.~He, S.~He, C.~Liu,
  W.~Shi, and S.~H. Bryant.
\newblock The {NCBI BioSystems} database.
\newblock {\em Nucleic Acids Research}, 38(S1):D492--D496, 2010.

\bibitem{pc-2011}
E.~G. Cerami, B.~E. Gross, E.~Demir, I.~Rodchenkov, {\"{O}}.~Babur, N.~Anwar,
  N.~Schultz, G.~D. Bader, and C.~Sander.
\newblock {Pathway Commons}, a web resource for biological pathway data.
\newblock {\em Nucleic Acids Research}, 39(S1):D685--D690, 2011.

\bibitem{cho-2006}
M.~Cho, J.~Kang, Y.~Moon, H.~Nam, et~al.
\newblock {STAT3} and {NF-kappaB} signal pathway is required for
  {IL-23-mediated} {IL-17} production in spontaneous arthritis animal model
  {IL-1} receptor antagonist-deficient mice.
\newblock {\em Journal of Immunology}, 176(9):5652--5661, 2006.

\bibitem{bonizzi-2004}
G.~Bonizzi and M.~Karin.
\newblock The two {NF-kappaB} activation pathways and their role in innate and
  adaptive immunity.
\newblock {\em Trends in Immunology}, 25(6):280--288, 2004.

\bibitem{charo-2006}
I.~F. Charo and R.~M. Ransohoff.
\newblock The many roles of chemokines and chemokine receptors in inflammation.
\newblock {\em New England Journal of Medicine}, 354(6):610--621, 2006.

\bibitem{dong-2002}
C.~Dong, R.~J. Davis, and R.~A. Flavell.
\newblock {MAP} kinases in the immune response.
\newblock {\em Annual Review of Immunology}, 20:55--72, 2002.

\bibitem{pao-2007}
L.~I. Pao, K.~Badour, K.~A. Siminovitch, and B.~G. Neel.
\newblock Nonreceptor protein-tyrosine phosphatases in immune cell signaling.
\newblock {\em Annual Review of Immunology}, 25:473--523, 2007.

\bibitem{oshea-2002}
J.~J. O'Shea, A.~Ma, and P.~Lipsky.
\newblock Cytokines and autoimmunity.
\newblock {\em Nature Reviews Immunology}, 2(1):37--45, 2002.

\bibitem{godessart-kunkel-2001}
N.~Godessart and S.~L. Kunkel.
\newblock Chemokines in autoimmune disease.
\newblock {\em Current Opinion in Immunology}, 13(6):670--675, 2001.

\bibitem{zhernakova-2009}
A.~Zhernakova, C.~C. van Diemen, and C.~Wijmenga.
\newblock Detecting shared pathogenesis from the shared genetics of
  immune-related diseases.
\newblock {\em Nature Reviews Genetics}, 10(1):43--55, 2009.

\bibitem{mcgovern-powrie-2007}
D.~McGovern and F.~Powrie.
\newblock The {IL23} axis plays a key role in the pathogenesis of {IBD}.
\newblock {\em Gut}, 56(10):1333--1336, 2007.

\bibitem{abraham-cho-2009}
C.~Abraham and J.~H. Cho.
\newblock {IL-23} and autoimmunity: new insights into the pathogenesis of
  inflammatory bowel disease.
\newblock {\em Annual Review of Medicine}, 60(1):97--110, 2009.

\bibitem{cho-2008}
J.~H. Cho.
\newblock The genetics and immunopathogenesis of inflammatory bowel disease.
\newblock {\em Nature Reviews Immunology}, 8(6):458--466, 2008.

\bibitem{hunter-2005}
C.~A. Hunter.
\newblock New {IL-12-family} members: {IL-23} and {IL-27}, cytokines with
  divergent functions.
\newblock {\em Nature Reviews Immunology}, 5(7):521--531, 2005.

\bibitem{hampe-2007}
J.~Hampe, A.~Franke, P.~Rosenstiel, A.~Till, et~al.
\newblock A genome-wide association scan of nonsynonymous {SNPs} identifies a
  susceptibility variant for {Crohn} disease in {ATG16L1}.
\newblock {\em Nature Genetics}, 39(2):207--211, 2007.

\bibitem{rioux-2007}
J.~D. Rioux, R.~J. Xavier, K.~D. Taylor, M.~S. Silverberg, et~al.
\newblock Genome-wide association study identifies new susceptibility loci for
  {Crohn} disease and implicates autophagy in disease pathogenesis.
\newblock {\em Nature Genetics}, 39(5):596--604, 2007.

\bibitem{homer-2010}
C.~R. Homer, A.~L. Richmond, N.~A. Rebert, J.~Achkar, and C.~McDonald.
\newblock {ATG16L1} and {NOD2} interact in an autophagy-dependent antibacterial
  pathway implicated in {Crohn's} disease pathogenesis.
\newblock {\em Gastroenterology}, 139(5):1630--1641, 2010.

\bibitem{go-2000}
M.~Ashburner, C.~A. Ball, J.~A. Blake, D.~Botstein, et~al.
\newblock {Gene Ontology}: tool for the unification of biology.
\newblock {\em Nature Genetics}, 25:25--29, 2000.

\bibitem{storey-2003}
J.~D. Storey.
\newblock The positive false discovery rate: a {Bayesian} interpretation and
  the q-value.
\newblock {\em Annals of Statistics}, 31(6):2013--2035, 2003.

\bibitem{storey-2003b}
J.~D. Storey and R.~Tibshirani.
\newblock Statistical significance for genomewide studies.
\newblock {\em Proceedings of the National Academy of Sciences},
  100(16):9440--9445, 2003.

\bibitem{mcvean-2004}
G.~A.~T. McVean, S.~R. Myers, S.~Hunt, P.~Deloukas, D.~R. Bentley, and
  P.~Donnelly.
\newblock The fine-scale structure of recombination rate variation in the human
  genome.
\newblock {\em Science}, 304(5670):581--584, 2004.

\bibitem{ucsc-genome-browser-2012}
T.~R. Dreszer, D.~Karolchik, A.~S. Zweig, A.~S. Hinrichs, et~al.
\newblock The {UCSC Genome Browser} database: extensions and updates 2011.
\newblock {\em Nucleic Acids Research}, 40:D918--D923, 2012.

\bibitem{fernando-2008}
M.~M.~A. Fernando, C.~R. Stevens, Ee.~C. Walsh, P.~L. De~Jager, P.~Goyette,
  R.~M. Plenge, T.~J. Vyse, and J.~D. Rioux.
\newblock Defining the role of the {MHC} in autoimmunity: A review and pooled
  analysis.
\newblock {\em PLoS Genetics}, 4(4):e1000024, 2008.

\bibitem{mathew-2008}
C.~G. Mathew.
\newblock New links to the pathogenesis of {Crohn} disease provided by
  genome-wide association scans.
\newblock {\em Nature Reviews Genetics}, 9(1):9--14, 2008.

\bibitem{rioux-2000}
J.~D. Rioux, M.~S. Silverberg, M.~J. Daly, A.~H. Steinhart, et~al.
\newblock Genomewide search in {Canadian} families with inflammatory bowel
  disease reveals two novel susceptibility loci.
\newblock {\em American Journal of Human Genetics}, 66(6):1863--1870, 2000.

\bibitem{silverberg-2007}
M.~S. Silverberg, R.~H. Duerr, S.~R. Brant, G.~Bromfield, et~al.
\newblock Refined genomic localization and ethnic differences observed for the
  {IBD5} association with {Crohn's} disease.
\newblock {\em European Journal of Human Genetics}, 15(3):328--335, 2007.

\bibitem{vanheel-2004}
D.~A. Van~Heel, S.~A. Fisher, A.~Kirby, M.~J. Daly, J.~D. Rioux, and C.~M.
  Lewis.
\newblock Inflammatory bowel disease susceptibility loci defined by genome scan
  meta-analysis of 1952 affected relative pairs.
\newblock {\em Human Molecular Genetics}, 13(7):763--770, 2004.

\bibitem{dejager-2009}
P.~L. {De Jager}, X.~Jia, J.~Wang, P.~I.~W. {de Bakker}, et~al.
\newblock Meta-analysis of genome scans and replication identify {CD6}, {IRF8}
  and {TNFRSF1A} as new multiple sclerosis susceptibility loci.
\newblock {\em Nature Genetics}, 41(7):776--782, 2009.

\bibitem{gorlova-2011}
O.~Gorlova, J.~Martin, B.~Rueda, B.~P.~C. Koeleman, et~al.
\newblock Identification of novel genetic markers associated with clinical
  phenotypes of systemic sclerosis through a genome-wide association strategy.
\newblock {\em PLoS Genetics}, 7(7):e1002178, 2011.

\bibitem{fisher-2008}
S.~A. Fisher, M.~Tremelling, C.~A. Anderson, R.~Gwilliam, et~al.
\newblock Genetic determinants of ulcerative colitis include the {ECM1} locus
  and five loci implicated in {Crohn's} disease.
\newblock {\em Nature Genetics}, 40(6):710--712, 2008.

\bibitem{franke-2008}
A.~Franke, T.~Balschun, T.~H. Karlsen, J.~Hedderich, et~al.
\newblock Replication of signals from recent studies of {Crohn's} disease
  identifies previously unknown disease loci for ulcerative colitis.
\newblock {\em Nature Genetics}, 40(6):713--715, 2008.

\bibitem{cargill-2007}
M.~Cargill, S.~J. Schrodi, M.~Chang, V.~E. Garcia, et~al.
\newblock A large-scale genetic association study confirms {IL12B} and leads to
  the identification of {IL23R} as psoriasis-risk genes.
\newblock {\em American Journal of Human Genetics}, 80(2):273--290, 2007.

\bibitem{hugot-2001}
J.~P. Hugot, M.~Chamaillard, H.~Zouali, S.~Lesage, et~al.
\newblock Association of {NOD2} leucine-rich repeat variants with
  susceptibility to {Crohn's} disease.
\newblock {\em Nature}, 411(6837):599--603, 2001.

\bibitem{cirulli-2010}
E.~T. Cirulli and D.~B. Goldstein.
\newblock Uncovering the roles of rare variants in common disease through
  whole-genome sequencing.
\newblock {\em Nature Reviews Genetics}, 11(6):415--425, 2010.

\bibitem{1000genomes}
{The 1000 Genomes Project Consortium}.
\newblock A map of human genome variation from population-scale sequencing.
\newblock {\em Nature}, 467(7319):1061--1073, 2010.

\bibitem{trynka-2011}
G.~Trynka, K.~A. Hunt, N.~A. Bockett, J.~Romanos, et~al.
\newblock Dense genotyping identifies and localizes multiple common and rare
  variant association signals in celiac disease.
\newblock {\em Nature Genetics}, 43(12):1193--1201, 2011.

\bibitem{bansal-2010}
V.~Bansal, O.~Libiger, A.~Torkamani, and N.~J. Schork.
\newblock Statistical analysis strategies for association studies involving
  rare variants.
\newblock {\em Nature Reviews Genetics}, 11(11):773--785, 2010.

\bibitem{neale-2011}
B.~M. Neale, M.~A. Rivas, B.~F. Voight, D.~Altshuler, B.~Devlin,
  M.~Orho-Melander, S.~Kathiresan, S.~M. Purcell, K.~Roeder, and M.~J. Daly.
\newblock Testing for an unusual distribution of rare variants.
\newblock {\em PLoS Genetics}, 7(3):e1001322, 2011.

\bibitem{guan-stephens-2008}
Y.~Guan and M.~Stephens.
\newblock Practical issues in imputation-based association mapping.
\newblock {\em PLoS Genetics}, 4(12):e1000279, 2008.

\bibitem{hapmap}
{International HapMap Consortium}.
\newblock A second generation human haplotype map of over 3.1 million {SNPs}.
\newblock {\em Nature}, 449(7164):851--861, 2007.

\bibitem{clayton-2005}
D.~G. Clayton, N.~M. Walker, D.~J. Smyth, R.~Pask, et~al.
\newblock Population structure, differential bias and genomic control in a
  large-scale, case-control association study.
\newblock {\em Nat Genetics}, 37(11):1243--1246, 2005.

\bibitem{price-2010}
A.~L. Price, N.~A. Zaitlen, D.~Reich, and N.~Patterson.
\newblock New approaches to population stratification in genome-wide
  association studies.
\newblock {\em Nature Reviews Genetics}, 11(7):459--463, 2010.

\bibitem{tintle-2011}
N.~Tintle, H.~Aschard, I.~Hu, N.~Nock, H.~Wang, and E.~Pugh.
\newblock Inflated type {I} error rates when using aggregation methods to
  analyze rare variants in the {1000 Genomes Project} exon sequencing data in
  unrelated individuals: summary results from {Group 7} at {Genetic Analysis
  Workshop 17}.
\newblock {\em Genetic Epidemiology}, 35(S1):S56--S60, 2011.

\bibitem{zeggini-2008}
E.~Zeggini, L.~J. Scott, R.~Saxena, B.~F. Voight, et~al.
\newblock Meta-analysis of genome-wide association data and large-scale
  replication identifies additional susceptibility loci for type 2 diabetes.
\newblock {\em Nature Genetics}, 40(5):638--645, 2008.

\bibitem{biocyc-2010}
R.~Caspi, T.~Altman, J.~M. Dale, K.~Dreher, et~al.
\newblock The {MetaCyc} database of metabolic pathways and enzymes and the
  {BioCyc} collection of pathway/genome databases.
\newblock {\em Nucleic Acids Research}, 38(S1):D473--D479, 2010.

\bibitem{humancyc-2004}
P.~Romero, J.~Wagg, M.~Green, D.~Kaiser, M.~Krummenacker, and P.~Karp.
\newblock Computational prediction of human metabolic pathways from the
  complete human genome.
\newblock {\em Genome Biology}, 6(1):R2, 2004.

\bibitem{kegg-2010}
M.~Kanehisa, S.~Goto, M.~Furumichi, M.~Tanabe, and M.~Hirakawa.
\newblock {KEGG} for representation and analysis of molecular networks
  involving diseases and drugs.
\newblock {\em Nucleic Acids Research}, 38(S1):D355--D360, 2010.

\bibitem{panther-2009}
H.~Mi and P.~Thomas.
\newblock {PANTHER Pathway}: an ontology-based pathway database coupled with
  data analysis tools.
\newblock In Y.~Nikolsky and J.~Bryant, editors, {\em Protein Networks and
  Pathway Analysis}, volume 563 of {\em Methods in Molecular Biology}, pages
  1230--140. 2009.

\bibitem{panther-2010}
H.~Mi, Q.~Dong, A.~Muruganujan, P.~Gaudet, Suzanna Lewis, and Paul~D. Thomas.
\newblock {PANTHER} version 7: improved phylogenetic trees, orthologs and
  collaboration with the {Gene Ontology Consortium}.
\newblock {\em Nucleic Acids Research}, 38(S1):D204--D210, 2010.

\bibitem{wikipathways-2011}
T.~Kelder, M.~P. van Iersel, K.~Hanspers, M.~Kutmon, B.~R. Conklin, C.~T.
  Evelo, and A.~R. Pico.
\newblock {WikiPathways}: building research communities on biological pathways.
\newblock {\em Nucleic Acids Research}, 40(D1):D1301--D1307, 2012.

\bibitem{wikipathways-2008}
A.~R. Pico, T.~Kelder, M.~P. van Iersel, K.~Hanspers, B.~R. Conklin, and
  C.~Evelo.
\newblock {WikiPathways}: pathway editing for the people.
\newblock {\em PLoS Biology}, 6(7):e184, 2008.

\bibitem{biopax-2010}
E.~Demir, M.~P. Cary, S.~Paley, K.~Fukuda, et~al.
\newblock The {BioPAX} community standard for pathway data sharing.
\newblock {\em Nature Biotechnology}, 28(9):935--942, 2010.

\bibitem{soh-2010}
D.~Soh, D.~Dong, Y.~Guo, and L.~Wong.
\newblock Consistency, comprehensiveness, and compatibility of pathway
  databases.
\newblock {\em BMC Bioinformatics}, 11(1):449, 2010.

\bibitem{stobbe-2011}
M.~Stobbe, S.~Houten, G.~Jansen, A.~van Kampen, and P.~Moerland.
\newblock Critical assessment of human metabolic pathway databases: a stepping
  stone for future integration.
\newblock {\em BMC Systems Biology}, 5(1):165, 2011.

\bibitem{cookson-2009}
W.~Cookson, L.~Liang, G.~Abecasis, M.~Moffatt, and M.~Lathrop.
\newblock Mapping complex disease traits with global gene expression.
\newblock {\em Nature Reviews Genetics}, 10(3):184--194, 2009.

\bibitem{dixon-2007}
A.~L. Dixon, L.~Liang, M.~F. Moffatt, W.~Chen, S.~Heath, K.~C.~C. Wong,
  J.~Taylor, E.~Burnett, I.~Gut, M.~Farrall, G.~M. Lathrop, G.~R. Abecasis, and
  W.~O.~C. Cookson.
\newblock A genome-wide association study of global gene expression.
\newblock {\em Nature Genetics}, 39(10):1202--1207, 2007.

\bibitem{stranger-2007}
B.~E. Stranger, A.~C. Nica, M.~S. Forrest, A.~Dimas, C.~P. Bird, C.~Beazley,
  C.~E. Ingle, M.~Dunning, P.~Flicek, D.~Koller, S.~Montgomery, S.~Tavare,
  P.~Deloukas, and E.~T. Dermitzakis.
\newblock Population genomics of human gene expression.
\newblock {\em Nature Genetics}, 39(10):1217--1224, 2007.

\bibitem{george-mcculloch-1993}
E.~I. George and R.~E. McCulloch.
\newblock Variable selection via {Gibbs} sampling.
\newblock {\em Journal of the American Statistical Association},
  88(423):881--889, 1993.

\bibitem{mitchell-beauchamp-1988}
T.~J. Mitchell and J.~J. Beauchamp.
\newblock Bayesian variable selection in linear regression.
\newblock {\em Journal of the American Statistical Association}, 83:1023--1032,
  1988.

\bibitem{ioannidis-2006}
J.~P.~A. Ioannidis, T.~A. Trikalinos, and M.~J. Khoury.
\newblock Implications of small effect sizes of individual genetic variants on
  the design and interpretation of genetic association studies of complex
  diseases.
\newblock {\em American Journal of Epidemiology}, 164(7):609--614, 2006.

\bibitem{obrien-2004}
S.~M. O'Brien and D.~B. Dunson.
\newblock Bayesian multivariate logistic regression.
\newblock {\em Biometrics}, 60(3):739--746, 2004.

\bibitem{jordan-1999}
M.~I. Jordan, Z.~Ghahramani, T.~S. Jaakkola, and L.~K. Saul.
\newblock An introduction to variational methods for graphical models.
\newblock {\em Machine Learning}, 37:183--233, 1999.

\bibitem{burden-faires}
R.~Burden and J.~D. Faires.
\newblock {\em Numerical analysis}.
\newblock Thomson Brooks/Cole, 2005.

\bibitem{bishop}
C.~M. Bishop.
\newblock {\em Pattern Recognition and Machine Learning}.
\newblock Springer, 2006.

\bibitem{jaakkola-jordan-2000}
T.~S. Jaakkola and M.~I. Jordan.
\newblock Bayesian parameter estimation via variational methods.
\newblock {\em Statistics and Computing}, 10:25--37, 2000.

\bibitem{oshea-2011}
J.~J. O'Shea, R.~Lahesmaa, G.~Vahedi, A.~Laurence, and Y.~Kanno.
\newblock Genomic views of {STAT} function in {CD4+ T} helper cell
  differentiation.
\newblock {\em Nature Reviews Immunology}, 11(4):239--250, 2011.

\bibitem{parham-2002}
C.~Parham, M.~Chirica, J.~Timans, E.~Vaisberg, M.~Travis, et~al.
\newblock A receptor for the heterodimeric cytokine {IL-23} is composed of
  {IL-12Rbeta1} and a novel cytokine receptor subunit, {IL-23R}.
\newblock {\em Journal of immunology}, 168(11):5699--5708, 2002.

\bibitem{rioux-2001}
J.~D. Rioux, M.~J. Daly, M.~S. Silverberg, K.~Lindblad, et~al.
\newblock Genetic variation in the {5q31} cytokine gene cluster confers
  susceptibility to crohn disease.
\newblock {\em Nature Genetics}, 29(2):223--228, 2001.

\bibitem{barrett-chandra-2011}
M.~Barrett and S.~Chandra.
\newblock A review of major {Crohn's} disease susceptibility genes and their
  role in disease pathogenesis.
\newblock {\em Genes and Genomics}, 33(4):317--325, 2011.

\bibitem{anderson-2011}
C.~A. Anderson, G.~Boucher, C.~W. Lees, A.~Franke, et~al.
\newblock Meta-analysis identifies 29 additional ulcerative colitis risk loci,
  increasing the number of confirmed associations to 47.
\newblock {\em Nature Genetics}, 43(3):246--252, 2011.

\bibitem{lees-2011}
C.~W. Lees, J.~C. Barrett, M.~Parkes, and J.~Satsangi.
\newblock New {IBD} genetics: common pathways with other diseases.
\newblock {\em Gut}, 60(12):1739--1753, 2011.

\end{thebibliography}

\appendix

\section{Supplementary Methods} 

\subsection*{Pathways}

We retrieve most of the pathways from the Pathway Commons
\cite{pc-2011} and NCBI BioSystems \cite{biosystems-2010}
repositories. From the Pathway Commons website, we download the
October 26, 2011 version of Gene Matrix Transposed {\tt (.gmt)} file
for {\em homo sapiens}. To retrieve BioSystems pathways, we first get
the pathway names and IDs by searching for {\tt ``homo
  sapiens''[organism]}, then save the search result as a CSV
file. Next, we download the November 15, 2011 version of the {\tt
  biosystems\_gene} file from the NCBI FTP site, which provides
associations between genes and pathways.  These two repositories
include pathways from the same databases, but due to differences in
versions of the databases and data processing procedure, there are
discrepancies among pathways. At present, BioSystems ignores nesting
relationships between pathways in the PID. This can lead to large
discrepancies in pathway gene sets, notably in the IL23-mediated
signaling pathway.\footnote{Personal communication with Lewis Geer and
  Emek Demir.} Since we cannot fully account for all discrepancies in
the BioSystems and Pathway Commons gene sets, whenever there is
disagreement we include both gene sets, and assess evidence for
enrichment of these gene sets separately in our analysis.

We download a version of the BioCarta database at
\url{www.openbioinformatics/gengen}. We use this version of the
BioCarta data because it was used in an previous pathway analysis of
Crohn's disease \cite{wang-2009}. We download version 3.01 of the
PANTHER ``sequence association'' file from their FTP site.
From the sequence association file, we retain lines containing {\tt
  ENSG*} accession numbers (corresponding to human genes) and
remove entries that do not map to Entrez gene IDs.

The numbers given in Table~\ref{table:pathways} are tabulated after
discarding 213 pathways with less than two genes that map to the
reference genome, and after removing 44 PID pathways from Pathway
Commons that contain over 500 genes because their definitions include
a large number of nested pathways. We include all groups of pathways
except for two unusually large gene sets from the KEGG database that
are unions of related pathways, ``metabolic pathways'' and ``pathways
in cancer.''

\subsection*{Computation}

The main difficulty in computing the Bayes factor (eq.~\ref{eq:BF}) is
the combinatorially large number of ways we can include SNPs in the
additive model of disease risk. In previous work \cite{bvs}, we
described an approximation that yields an efficient procedure for
computing the likelihood and PIPs. (Actually, this approximation was
derived for the specific case when all variables have the same prior
inclusion probability, or when $\theta = 0$, but it is straightforward
to extend this approximation to the more general case with $\theta >
0$.) Once we have a recipe for efficiently computing $p(y \,|\,
\mathbf{X}, a, \theta_0, \theta)$, we are left with the task of
computing a one-dimensional integral in the denominator of
\eqref{eq:BF-2}, and a double integral in the numerator. Each of these
integrals is then approximated using simple numerical integration
techniques.

The basic idea behind this approximation is to formulate a lower
bound to the likelihood,
\begin{align}
p(y \,|\, \mathbf{X}, a, \theta_0, \theta) \geq 
e^{F(\Data, \theta_0, \theta, \phi)},
\label{eq:lower-bound-intro}
\end{align}
then to adjust the free parameters, denoted by $\phi$, so that this
bound is as tight as possible. (The exact form of $F(\Data, \theta_0,
\theta, \phi)$ is derived in \cite{bvs}, and is given below.) This
lower bound is formulated by introducing a probability distribution
$q(\beta; \phi)$ that approximates the posterior of $\beta$ given
$\theta_0$ and $\theta$ so that maximizing the lower bound corresponds
to finding the approximating distribution that best matches the
posterior. More precisely, it amounts to searching for the free
parameters $\phi$ that minimize the Kullback-Leibler divergence
between $q(\beta; \phi)$ and the posterior of $\beta$ given $\theta_0$
and $\theta$ \cite{jordan-1999}. The trick to making this approach
tractable lies in forcing $q(\beta; \phi)$ to observe a simple
conditional independence property, as originally suggested by
\cite{logsdon}: each regression coefficient $\beta_j$ is independent
of the other coefficients {\em a posteriori} given $\theta_0$ and
$\theta$. In other words, we restrict this distribution to be of the
form
\begin{align}
\textstyle q(\beta; \phi) = \prod_{j=1}^p q(\beta_j; \phi_j), 
\label{eq:q}
\end{align}
where $\phi_j$ is the set of free parameters corresponding to the
$j$th factor.

For most SNPs, this conditional independence assumption is
appropriate---most SNPs are unlinked because they are on separate
chromosomes, or they are weakly linked because of recombination. In
this case, the fully-factorized approximation $q(\beta; \phi)$ will
closely recover the correct posterior distribution of the additive
effects for these SNPs. But the conditional independence assumption is
violated for SNPs in linkage disequilibrium. In that case, we do not
expect to obtain accurate posterior statistics and, in practice, we
find that the lower bound \eqref{eq:lower-bound-intro} can be a poor
substitute to the correct likelihood. However, we are interested in
accurate computation of BFs, not individual likelihoods, so what
matters is whether $e^{F(\Data, \theta_0, \theta, \phi)}$ correctly
captures the shape of the likelihood, or how the likelihood $p(y \,|\,
\mathbf{X}, a, \theta_0, \theta)$ changes as a function of $\theta_0$
and $\theta$; if the lower bound undershoots the exact likelihood by a
constant factor across different settings of $\theta_0$ and $\theta$,
this constant factor will cancel out in the BF. In \cite{bvs}, we show
that the variational approximation (when the phenotype is a
quantitative trait) can closely reproduce the shape of the likelihood,
and can give accurate estimates of some posterior quantities, even
when the conditional independence assumptions are not particularly
appropriate.  We caution, however, that the accuracy the approximation
has only been assessed empirically, and we have no theoretical
guarantees of its accuracy.

\subsubsection*{Computing the Bayes factor for one pathway}

To start, we formulate a simple piecewise numerical approximation to
the integrals in \eqref{eq:BF-2} based on Simpson's rule
\cite{burden-faires}. We replace each instance of the likelihood $p(y
\,|\, \mathbf{X}, a, \theta_0, \theta)$ with its corresponding lower
bound~\eqref{eq:lower-bound-intro}. Following the discussion above, we
have
\begin{align}
\BF(a) &\approx
\frac{\iint e^{F(\Data, \theta_0, \theta, \phi(\theta_0,\theta))} \, 
            p(\theta_0) \, p(\theta) \, d\theta \, d\theta_0}
           {\int e^{F(\Data, \theta_0, \theta=0, 
            \phi(\theta_0,\theta=0))} \, p(\theta_0) \, d\theta_0},
\end{align}
so the numerical approximation to the BF is
\begin{align}
&\BF(a) \approx
\frac{I_{\rm alt}}{I_{\rm null}} \nonumber \\ 
&= \frac{\sum_i \sum_j w_{ij} 
               e^{F(\Data, \theta_0^{(i)}, \theta^{(j)}, 
               \phi(\theta_0^{(i)},\theta^{(j)}))} \, 
            p(\theta_0^{(i)}) \, p(\theta^{(j)})}
           {\sum_i w_i e^{F(\Data, \theta_0^{(i)}, \theta=0, 
            \phi(\theta_0^{(i)},\theta=0))} \, p(\theta_0^{(i)})},
\label{eq:BF-approx}
\end{align}
where $w_i$ and $w_{ij}$ are weights obtained by applying Simpson's
rule. To calculate the numerical approximation to the null likelihood,
$I_{\rm null}$, we evaluate the lower bound at equally spaced points
over the interval $[-6,-2]$. The likelihood under the alternative is a
double integral, so we evaluate the lower bound at points on a regular
grid over the rectangular region $\theta_0 \in [-6,-2]$, $\theta \in
[0,3]$. In eq.~\ref{eq:BF-approx}, the free parameters $\phi$ are
expressed as a function of $\theta_0$ and $\theta$ because we adjust
them separately for each setting of the hyperparameters $(\theta_0,
\theta)$. (Optimizing $\phi$ involves iterating coordinate ascent
steps until these steps converge to stationary point which constitutes
a locally optimal bound to the likelihood. Full details about the
procedure to solve for $\phi$ are given in \cite{bvs}. This procedure
scales linearly with the number of samples and the number of SNPs.)
Adjusting the free parameters for each setting of the hyperparameters
can be an costly endeavor for a large problem, so to reduce the
expense of computing the BF we formulate piecewise numerical
approximations to the integrals using a small number of equally spaced
points at intervals of length $0.25$, so $\theta_0^{(i)} = -6, -5.75,
\ldots, -2$ and $\theta^{(j)} = 0, 0.25, \ldots, 3$. This coarse
partitioning risks some loss of accuracy, especially if the posterior
distribution of the hyperparameters is sharply peaked inside the
subintervals, and a finer grid is certainly possible. An adaptive
method that refines the subintervals in the piecewise approximation
could have been used instead \cite{burden-faires}, but we stick to
this simple scheme with equally spaced points at larger subintervals
because it allows us to compute BFs for all $\about{3000}$ candidate
pathways in a reasonable amount of time.

The analytic expression for the lower bound to the log-likelihood is
derived in the Appendix of \cite{bvs} and, for convenience, we
reproduce it here:
\begin{align}
F(\Data, \theta_0, \theta, \phi)
=\,& \log\hat{\sigma}_0 + \frac{\bar{y}^2}{2\bar{u}} + 
\sum_{i=1}^n \log\psi(\eta_i) 
\nonumber \\ 
& + \smfrac{\eta_i}{2}(u_i\eta_i - 1)
+ \hat{y}^T\mathbf{X}r - \half r^T\mathbf{X}^T\hat{U}\mathbf{X}r
\nonumber \\
& 
- \frac{1}{2} \sum_{j=1}^p (\mathbf{X}^T\hat{U}\mathbf{X})_{jj} \Var\lb\beta_j\rb
\nonumber \\ 
& + \sum_{j=1}^p \frac{\alpha_j}{2} \bigg[
1 + \log\bigg(\frac{s_j^2}{\sigma_a^2}\bigg)
  - \frac{s_j^2 + \mu_j^2}{\sigma_a^2} \bigg]
\nonumber \\
& 
- \sum_{j=1}^p \alpha_j \log\Big(\frac{\alpha_j}{\pi_j}\Big)
\nonumber \\
& - \sum_{j=1}^p (1-\alpha_j) \log\Big(\frac{1-\alpha_j}{1-\pi_j}\Big),
\label{eq:lower-bound}
\end{align}
where the sigmoid function $\psi(x)$ is defined above, and the prior
inclusion probability $\pi_j$ for each SNP is given by
\eqref{eq:pathway-prior}.  For this expression we introduce the
following definitions: $\alpha_j$ is the PIP for SNP $j$ with respect
to the approximating distribution $q(\beta; \phi)$, $\mu_j$ and
$s_j^2$ are the approximate mean and variance of coefficient $\beta_j$
conditioned on being included in the model, $\Var\lb\beta_j\rb =
\alpha_j(\mu_j^2 + s_j^2) - (\alpha_j\mu_j)^2$ is the variance of
$\beta_j$ with respect to the approximating distribution, $r$ is a
column vector with entries $r_j = \alpha_j \mu_j$, $\hat{\sigma}_0 =
1/\sqrt{\bar{u}}$ is the standard deviation of the intercept $\beta_0$
given $\beta$, $\hat{\beta}_0 = \bar{y}/\bar{u}$ is the posterior mode
of the intercept $\beta_0$ when $\beta = 0$,
$(\mathbf{X}^T\hat{U}\mathbf{X})_{jj}$ is the $j$th diagonal entry of
matrix product $\mathbf{X}^T\hat{U}\mathbf{X}$, and we define $\bar{u}
= \sum_{i=1}^n u_i$, $\bar{y} = \sum_{i=1}^n (y_i - \frac{1}{2})$,
$\hat{y} = y - \frac{1}{2} - \hat{\beta}_0 u$, $\hat{U} = U -
uu^T/\bar{u}$, $u$ is a column vector with entries $u_i =
(\psi(\eta_i) - \frac{1}{2})/\eta_i$, and $U$ is the $n \times n$
matrix with diagonal entries $u_i$.

In \cite{bvs}, to derive this analytic expression for the lower bound 
we made an additional approximation to the nonlinear factors appearing
in the logistic regression likelihood $p(y \,|\, \mathbf{X}, \beta_0,
\beta)$, following \cite{bishop,jaakkola-jordan-2000}. This
approximation introduces an additional set of free parameters, $\eta =
(\eta_1, \ldots, \eta_n)$, so implicitly the lower bound
\eqref{eq:lower-bound} is a function of $\eta$ as well. Like $\phi$,
we adjust $\eta$ separately for each hyperparameter setting
$(\theta_0, \theta)$. The procedure to solve for $\eta$ is given in
\cite{bvs}.

Since the coordinate ascent updates used to solve for $\phi$ and
$\eta$ are only guaranteed to converge to a local maximum of the lower
bound, the choice of starting point can affect the tightness of the
lower bound, and the quality of the approximation. As we explain in
\cite{bvs}, this issue can be addressed somewhat by using a common
initialization $(\phi^{\rm (init)}, \eta^{\rm (init)})$ for the
coordinate ascent updates across all grid points $(\theta_0^{(i)},
\theta^{(j)})$, in which this initialization is selected by first
running the coordinate ascent procedure separately for each grid
point, with random initializations for $\phi$ and $\eta$, then
assigning $(\phi^{\rm (init)}, \eta^{\rm (init)})$ to the solution
from the hyperparameter setting with the largest marginal likelihood
estimate. In practice, when we follow this procedure we find that
final estimates of BFs and posterior statistics vary only slightly
when the analysis is re-run with different random initializations.
However, we cannot guarantee the possibility that a new random
starting point leads to the discovery of a much better approximation.

\subsubsection*{Computing the posterior inclusion probabilities and other 
posterior statistics}

In this section, we describe computation of PIPs and other posterior
statistics when a pathway, or a combination of pathways, is enriched.
Computation of these quantities under the null hypothesis proceeds in
a similar manner by setting $\theta = 0$.

Following the procedure for computing the BFs, we formulate a
piecewise numerical approximation to the integral in \eqref{eq:PIP},
substituting each PIP conditioned on $\theta_0$ and $\theta$ with the
corresponding variational approximation, $\alpha_k(\theta_0,\theta)
\approx p(\beta_k \neq 0 \,|\, \Data, \theta_0, \theta)$. This yields
the following approximate PIP:
\begin{align}
\PIP(k) \approx \textstyle \sum_i \sum_j 
        \tilde{w}_{ij} \, \alpha_k(\theta_0^{(i)}, \theta^{(j)}),
\label{eq:PIP-approx}
\end{align}
where we define
\begin{align*}
\tilde{w}_{ij} \propto
w_{ij} \, e^{F(\Data, \theta_0^{(i)}, \theta^{(j)}, 
      \phi(\theta_0^{(i)},\theta^{(j)}))} \, 
      p(\theta_0^{(i)}) \, p(\theta^{(j)}),
\end{align*}
such that $\sum_i \sum_j \tilde{w}_{ij} = 1$. Other posterior
quantities are computed by averaging over $\theta_0$ and $\theta$ in a
similar way. For example, the posterior mean enrichment estimate is
\begin{align}
\bar{\theta} &= E\lb \theta \,|\, \Data \rb \nonumber \\
&= \textstyle \iint \theta \, p(\theta_0, \theta \,|\, \Data) \, 
d\theta_0 \, d\theta
\approx \sum_i \sum_j \tilde{w}_{ij} \, \theta^{(j)}.
\label{eq:post-mean-theta}
\end{align}

To compute credible intervals for $\theta$, we add up the normalized
weights $\tilde{w}_{ij}$ over successively wider intervals of
$\theta$, beginning at the posterior mean $\bar{\theta}$, until the
sum of the normalized weights $\tilde{w}_{ij}$ reaches 0.95. As a
result, the credible intervals are at the same resolution as the grid
points used for the numerical approximation.

The estimate of the posterior probability that at least one SNPs in a
given segment of the genome is included in the additive model of
disease risk is
\begin{align}
P_1 &\approx \textstyle \sum_i \sum_j \tilde{w}_{ij} \, 
p(S \geq 1 \,|\, \Data, \theta_0^{(i)}, \theta^{(j)})
\nonumber \\
&= \textstyle \sum_i \sum_j \tilde{w}_{ij} 
\big[1 - p(S = 0 \,|\, \Data, \theta_0^{(i)}, \theta^{(j)})\big].
\label{eq:P1}
\end{align}
where $S = n$ represents the event that $n$ SNPs in the segment are
included in the model. Assume without loss of generality that SNPs in
the segment are labeled 1 through $m$. Since the regression
coefficients are independent under the fully-factorized approximating
distribution given $\theta_0$ and $\theta$, we have
\begin{align}
p(S = 0 \,|\, \Data, \theta_0, \theta) 
&= p(\beta_1 = 0 \wedge \cdots \wedge \beta_m = 0 
\,|\, \Data, \theta_0, \theta) \nonumber \\
&\approx \textstyle \prod_{k=1}^m (1-\alpha_k(\theta_0,\theta)), 
\label{eq:P1-approx-identity}
\end{align}
so our final estimate of this posterior statistic is
\begin{align}
P_1 \approx \textstyle \sum_i \sum_j \tilde{w}_{ij}
\Big[1 - \prod_{k=1}^m (1-\alpha_k(\theta_0^{(i)},\theta^{(j)}))\Big].
\label{eq:P1-approx}
\end{align}
To compute $P_2 = p(S \geq 2 \,|\, \Data)$ for a given segment, we
observe that $p(S \geq 2) = 1 - p(S = 1) - p(S = 0)$, and under the
fully-factorized variational approximation, we have that
\begin{align}
&p(S = 1 \,|\, \Data, \theta_0, \theta) \nonumber \\
&\quad =
p(\beta_1 \neq 0 \wedge \beta_2 = 0 \wedge \cdots \wedge \beta_m = 0
  \,|\, \Data, \theta_0, \theta) + \cdots \nonumber \\
& \qquad + p(\beta_1 = 0 \wedge \cdots \wedge \beta_{m-1} = 0 
            \wedge \beta_m \neq 0 \,|\, \Data, \theta_0, \theta) 
  \nonumber \\
&\quad\approx \Bigg[\prod_{k=1}^m (1-\alpha_k) \Bigg]
         \times \Bigg[ \sum_{k=1}^m \frac{\alpha_k}{1-\alpha_k} \Bigg].
\label{eq:P2-approx-identity}
\end{align}

\subsubsection*{Scaling computation to many pathways}

Numerical integration together with the variational approximation
makes it practical to compute $\BF(a)$ for one pathway, but computing
BFs for thousands of pathways is still a costly undertaking. We
introduce a simplifying assumption which, we show, yields substantial
savings in computation. We make the assumption that SNPs outside the
enriched pathway are unaffected by the pathway enrichment {\em a
  posteriori}. Formally, this means that $p(\beta_{\bar{A}} \,|\,
\Data, \theta_0, \theta) = p(\beta_{\bar{A}} \,|\, \Data, \theta_0,
\theta = 0)$, where $A$ is the set of SNPs assigned to the enriched
pathway, and $\bar{A}$ is the remaining set of SNPs. In other words,
the posterior distribution of the regression coefficients for SNPs
outside the enriched pathway remains the same under the null and
enrichment models. With this assumption, the posterior distribution of
$\beta$ given the hyperparameters $\theta_0$ and $\theta$ becomes
\begin{align}
p(\beta \,|\, \Data, \theta_0, \theta) &= 
p(\beta_A \,|\, \Data, \theta_0, \theta, \beta_{\bar{A}}) 
\nonumber \\ 
& \qquad \times p(\beta_{\bar{A}} \,|\, \Data, \theta_0, \theta = 0).
\label{eq:assumption}
\end{align}
This assumption amounts to conditioning on the additive effects of
SNPs outside the enriched pathway.

It is of course possible that SNPs contributing evidence for pathway
enrichment are correlated with SNPs outside the pathway, invalidating
this assumption. But because we assign SNPs to pathways in contiguous
blocks (we annotate all SNPs within 100 kb of a gene in the pathway;
see Methods), and because the way we assign SNPs to genes is not
precise (many SNPs assigned to a gene are probably not relevant to the
gene), errors as a result of this assumption are expected to be minor
compared to the imprecision in the SNP-pathway assignments. On the
other hand, if we were to assign SNPs to pathways more precisely
(e.g. a SNP known to modulate expression of a gene in the pathway),
then this assumption would not be appropriate.

\begin{figure*}[t]
\centering
\includegraphics[width=6in,keepaspectratio=true]{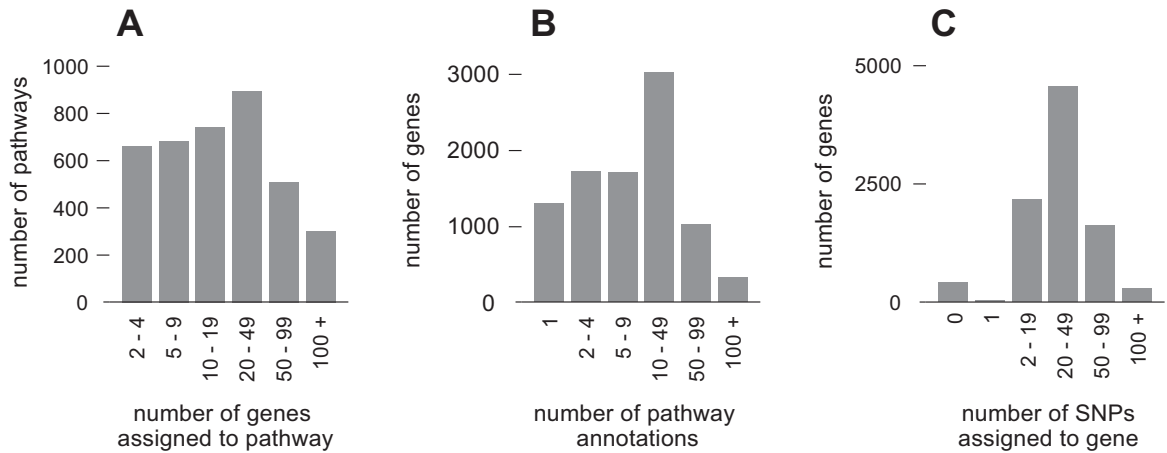}
\caption{{\em Panel A:} histogram of gene set sizes for pathways used
  in the analysis. {\em Panel B:} histogram of the number of pathways
  assigned to genes. {\em Panel C:} histogram of the number of SNPs
  assigned to genes. These counts include multiple versions of
  pathways from Pathway Commons and BioSystems.}
\label{fig:pathways} 
\end{figure*}

Next, we show how this assumption allows us to reuse computations.
This assumption implies that, for any SNP $j$ that is not assigned to
the enriched pathway, $q(\beta_j; \phi_j)$ remains the same under the
null and enrichment models; that is, for any $j \notin A$, $\phi_j =
\phi_j^{\ast}$, in which $q(\beta; \phi)$ approximates the posterior
distribution of $\beta$ given $\theta_0$, and $\theta = 0$, and
$q(\beta; \phi^{\ast})$ approximates the posterior distribution of
$\beta$ given $\theta_0$ and $\theta > 0$. From this result, the lower
bound can be written as
\begin{align}
&F(\Data, \theta_0, \theta, \phi^{\ast}) = 
F(\Data, \theta_0, \theta = 0, \phi) \nonumber \\
& \qquad
+ F(\{\mathbf{X}_A, \hat{y}_A, a_A = \mathbf{1}\}, \theta_0, \theta, 
    \phi_A^{\ast})
\nonumber \\ & \qquad
- F(\{\mathbf{X}_A, \hat{y}_A, a_A = \mathbf{1}\}, \theta_0, \theta = 0, 
    \phi_A^{\ast}),
\label{eq:lower-bound-with-assumption}
\end{align}
where $\mathbf{X}_A$ is the matrix of genotypes for SNPs assigned to the
enriched pathway, $\phi_A$ is the set of free parameters corresponding
to SNPs $j \in A$, $a_A$ is the set of pathway annotations restricted
to SNPs $j \in A$ (so $a_A$ is a vector of ones), and $\hat{y}_A = y -
\hat{U} \mathbf{X}_{\bar{A}} r_{\bar{A}}$ is the vector of binary labels
``corrected'' for SNPs outside the pathway, where $\mathbf{X}_{\bar{A}}$
is the genotype matrix for SNPs $j \notin A$, and $r_{\bar{A}}$ is the
vector $r$ with entries $r_j = \alpha_j \mu_j$ restricted to SNPs $j
\notin A$. Note that \eqref{eq:lower-bound-with-assumption} is valid
only if $\eta$ is held constant.

Identity \eqref{eq:lower-bound-with-assumption} suggests a way to
reuse our computations: once we have solved for
$\phi(\theta_0,\theta=0)$, the free parameters $\phi$ that (locally)
maximize the lower bound under the null hypothesis ($\theta=0$) for a
given $\theta_0$, to solve for $\phi(\theta_0, \theta)$ for any
$\theta > 0$, we only need to adjust the free parameters $\phi_j$
corresponding to SNPs $j$ in the enriched pathway. Crucially, $\eta$
must be held constant in \eqref{eq:lower-bound-with-assumption}, so
for any $\theta>0$, we set $\eta(\theta_0,\theta)$, the free
parameters $\eta$ adjusted for setting $(\theta_0,\theta)$ of the
hyperparameters, to $\eta(\theta_0,\theta=0)$.

\section{Supplementary Results}

\subsection*{Pathway database statistics}

\begin{figure*}[t]
\centering
\includegraphics[width=\textwidth,keepaspectratio=true]{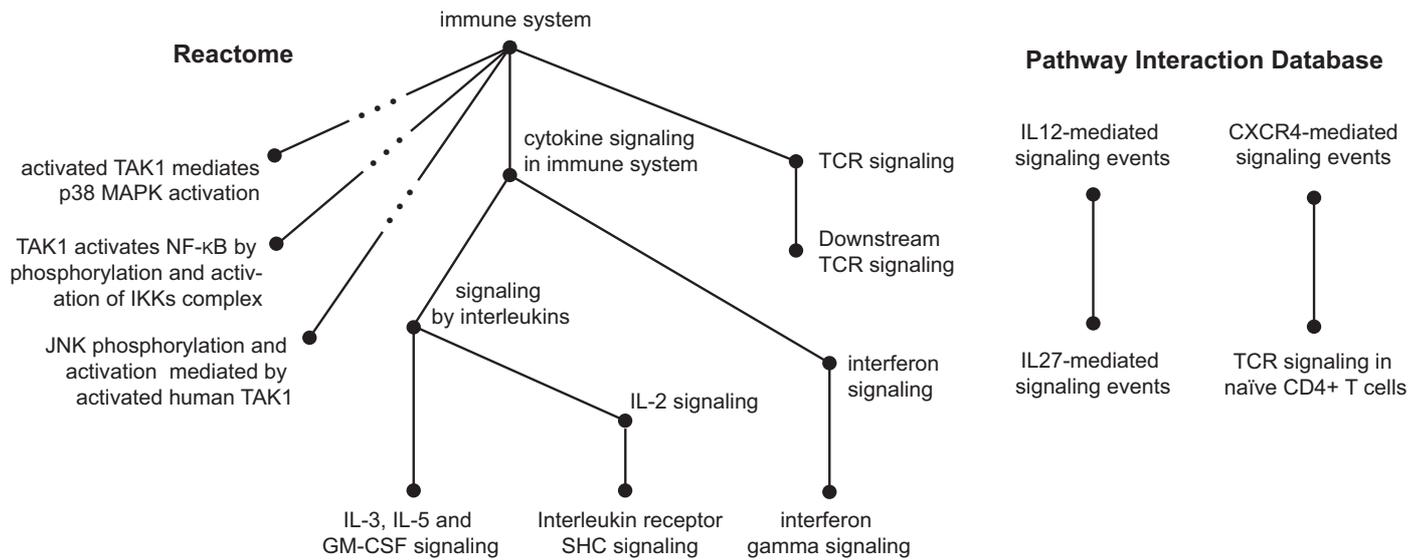}
\caption{Hierarchical relationships among Reactome and PID pathways
  listed in Table~\ref{table:ranking} and elsewhere in our analysis.
  For example, ``signaling by interleukins'' and ``interferon
  signaling'' (in addition to other pathways not shown) are part of
  ``cytokine signaling in immune system.''  Ellipses ($\cdots$)
  indicate that there are intermediate entries in the hierarchy not
  shown.}
\label{fig:hierarchy} 
\end{figure*}

We observe a wide range in the number of genes assigned to each
pathway (Fig.~\ref{fig:pathways}A). Some of the larger gene sets are
groups of related pathways in the Reactome and PID hierarchies. Out of
${\about}$23,000 genes in the reference genome, 9054 (${\about}39\%$)
are assigned to at least one pathway. The number of pathway
assignments per gene varies widely (Fig.~\ref{fig:pathways}B). Among
genes assigned to at least one pathway, 95\% are within 100 kb of a
SNP, and 45\% of SNPs are mapped to at least one gene in a pathway
(Fig.~\ref{fig:pathways}C). Figure~\ref{fig:hierarchy} depicts the
hierarchical relationships among pathways in
Table~\ref{table:ranking}.

\subsection*{More details on associations given enrichment 
of cytokine signaling genes}

Here we supply a few more details about regions of the genome that
show strong support for association only after accounting for
enrichment of cytokine signaling: the MHC class II region, the {\sf\em
  IBD5} locus, and a region at 17q21 near gene {\sf\em STAT3}. Without
pathway enrichment, the segment with the highest $P_1$ at 17q21 is
only $0.12$, but this increases to $0.79$ in the model in which
cytokine signaling is enriched. This association was first identified
in a meta-analysis \cite{barrett-2008}, and was later confirmed in
\cite{franke-2010}.  {\sf\em STAT3}, the most compelling
disease-susceptibility gene at this locus, plays a key role in Th17
cell differentiation and IL-23 signaling \cite{abraham-cho-2009,
  oshea-2011, parham-2002}. It has also been identified as a risk
factor for other autoimmune diseases, including ulcerative colitis
\cite{fisher-2008, franke-2008}. 

Enrichment of cytokine signaling also yields greater support for an
association in the {\sf\em IBD5} region at position 5q31, increasing
the probability of an included SNP from $P_1 = 0.18$ to $P_1 = 0.81$.
The {\sf\em IBD5} locus was first identified in a genome-wide study of
individuals from Quebec \cite{rioux-2000}, and this finding has since
been replicated and refined by several studies \cite{barrett-2008,
  franke-2010, rioux-2001, silverberg-2007}. It was identified as a
modest association for Crohn's disease in the original report (Table 4
in \cite{wtccc}) with trend $p$-value $5.4 \times 10^{-6}$ and $\BF =
10^{4.54}$. The {\sf\em IBD5} locus includes several candidate genes
in a region of extensive linkage disequilibrium. Despite this,
identification of this locus has suggested the contribution of gene
variants affecting epithelial barrier integrity in Crohn's disease
pathogenesis \cite{barrett-chandra-2011, vanlimbergen-2009}.

The MHC class II region, previously identified as a region showing
moderate evidence of association \cite{wtccc}, becomes a stronger
association when cytokine signaling is enriched, as $P_1$ increases
from 0.49 to 0.98. The MHC is one of the most thoroughly studied
regions of the human genome for its contribution to regulation of the
immune system. Association of this locus (also known as {\sf\em IBD3})
is widely replicated for Crohn's disease \cite{fernando-2008,
  mathew-2008, vanheel-2004, franke-2010}, ulcerative colitis
\cite{anderson-2011} and, not surprisingly, many other autoimmune
diseases \cite{fernando-2008, lees-2011}. The extensive linkage
disequilibrium across this gene-dense region complicates
identification of disease-susceptibility variants; see
\cite{fernando-2008} for a detailed examination of specific MHC genes
contributing to Crohn's disease risk. We also note that evidence for
association of the MHC class I genes, including {\sf\em TNF}, is low
under the null in our analysis, with $P_1 = 0.08$, and increases to
$P_1 = 0.52$ once we account for enrichment of cytokine signaling.

\subsection*{Additional results of sensitivity analysis}

\begin{figure*}[t]
\begin{center}
\includegraphics[width=2.3in,keepaspectratio=true]{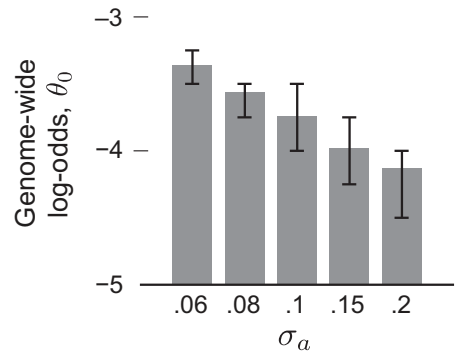}
\caption{Posterior mean of $\theta_0$ for each choice of
  $\sigma_a$. Error bars depict 95\% credible intervals.}
\label{fig:sensitivity-of-theta0} 
\end{center}
\end{figure*}

\begin{figure*}[t]
\begin{center}
\includegraphics[width=\textwidth,keepaspectratio=true]
{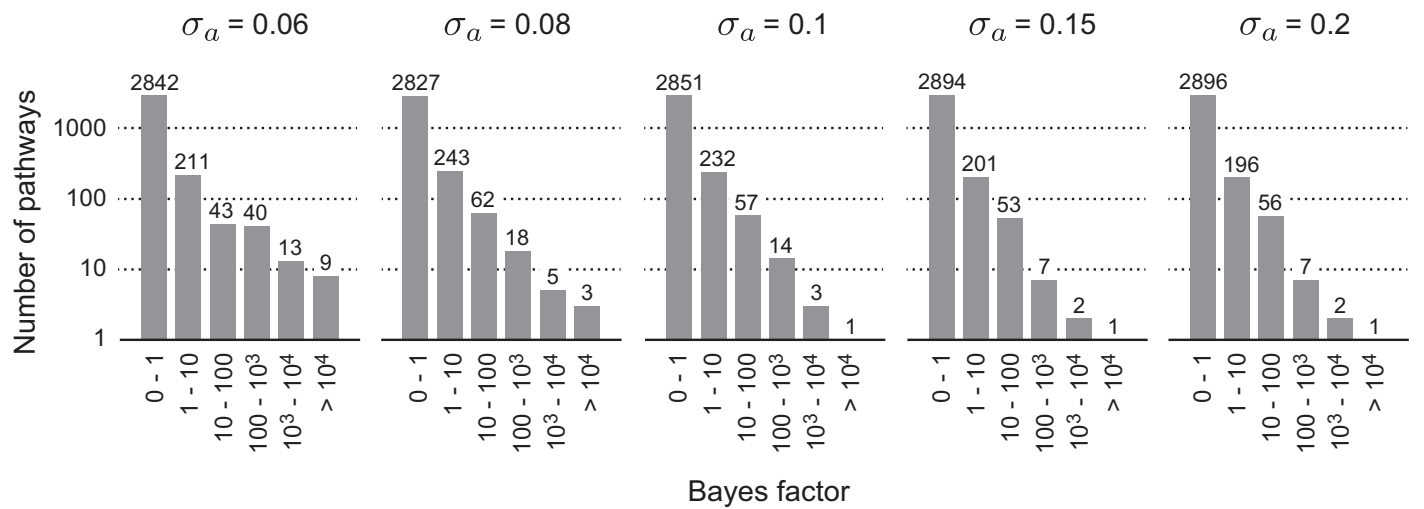}
\caption{Distribution of BFs for different settings of $\sigma_a$.}
\label{fig:sensitivity-of-bayes-factors} 
\end{center}
\end{figure*}

Under the null hypothesis, there is a clear trend in the overall
effect of the choice of $\sigma_a$ on the distribution of
associations; we observe that the posterior mean of the genome-wide
log-odds $\theta_0$ increases as $\sigma_a$ decreases
(Fig.~\ref{fig:sensitivity-of-theta0}). For instance, the normal prior
with standard deviation 0.06 corresponds to a posterior mean of
$\bar{\theta}_0 = -3.4$ and prior inclusion probabilities of roughly
$10^{-3.4} \approx 0.00044$ under the null hypothesis, which is more
than double the proportion of SNPs that are independent associations
(0.00018) when $\sigma_a = 0.1$.

Fig.~\ref{fig:sensitivity-of-bayes-factors} depicts the distribution
of BFs for different choices of $\sigma_a$. For most of the candidate
pathways, the support for enrichment does not change much.

\end{document}